\UseRawInputEncoding
\documentclass[epj]{svjour}
\usepackage{graphicx}
\usepackage{dcolumn}% Align table columns on decimal point
\usepackage[export]{adjustbox}
\graphicspath{ {Figures/} }
\usepackage{fnpct}
\usepackage[colorlinks]{hyperref}
\usepackage{amsmath}
\usepackage{amsxtra}
\usepackage{amstext}
\usepackage{amssymb}
\usepackage{latexsym}
\usepackage{dsfont}
\usepackage{textcomp}
\usepackage{ulem}
\usepackage{siunitx}
\makeatletter
\def\cl@chapter{\@elt {theorem}}
\makeatother
\usepackage[capitalise]{cleveref}
\usepackage{multirow}
\usepackage{color}
\usepackage{caption, subcaption}
\usepackage{bm}
\usepackage{soul}
\usepackage[square,numbers,sort&compress]{natbib}
\usepackage{hyperref}
\usepackage[english]{babel}
\usepackage{lipsum}
\usepackage{commath}
\usepackage[toc,page]{appendix}
\usepackage{dcolumn}% Align table columns on decimal point
\usepackage{mathtools,upgreek}
\usepackage{tabularx,ragged2e,booktabs,caption}
\usepackage{bigdelim,bigstrut}
\newcolumntype{C}[1]{>{\Centering}m{#1}}

\newcommand{\vect}[1]{\bm{#1}}
\newcommand{\etal}{\textit{et~al}. }
\newcommand{\stkout}[1]{\ifmmode\text{\sout{\ensuremath{#1}}}\else\sout{#1}\fi}
\hypersetup{
	colorlinks=true,
	linkcolor=blue,
	filecolor=blue,      
	urlcolor=blue,
}

\begin{document}
\newcommand{\bluu}[1]{\textcolor{blue}{#1}} %for displaying blu texts
\newcommand{\gre}[1]{\textcolor[rgb]{0,0.45,0}{#1}} %for displaying green texts
\newcommand{\rood}[1]{\textcolor{red}{[#1]}} %for displaying red texts
\newcommand{\pdx}[1][]{\pd{x}#1}
\newcommand{\pdt}[1][]{\pd{t}#1}
\newcommand{\eps}{\varepsilon}
\newcommand{\sig}{\sigma}
\newcommand{\kap}{\kappa}
\newcommand{\gam}{\gamma}
\newcommand{\om}{\omega}
\newcommand{\bbet}{\bm{\beta}}
\newcommand{\bkap}{\bm{\kap}}
\newcommand{\Vol}{\mathcal{V}}
\newcommand{\Sur}{\mathcal{S}}
\newcommand{\area}{\mathcal{A}}
\newcommand{\TM}{\mathsf{T}}
\newcommand{\SM}{\mathsf{S}}
\newcommand{\TSM}{\Tilde{\SM}}
\newcommand{\Oh}{\mathcal{O}}
\newcommand{\pt}{\mathcal{PT}}
\newcommand{\ii}{\mathrm{i}}
\newcommand{\bLozenge}{\mathbin{\blacklozenge}}

\title{Exceptional points and scattering of discrete mechanical metamaterials}
%\subtitle{Do you have a subtitle?\\ If so, write it here}
\author{Weidi Wang \and Alireza V. Amirkhizi
\thanks{\emph{email:} alireza\_amirkhizi@uml.edu}%
}% etc
% \thanks is optional - remove next line if not needed

\institute{Department of Mechanical Engineering, University of Massachusetts, Lowell, 
%\and
	Lowell, Massachusetts 01854, USA}
\date{Received: date / Revised version: date}
% The correct dates will be entered by Springer
%
\abstract{
Exceptional points (EPs) are complex singularities of parametric linear operators where two or more eigenvalues and eigenvectors coalesce. EPs are attracting increasing interest in mechanical metamaterials due to their strong potentials for wave filtering, cloaking, and sensing applications. 
This work studies the band topology and scattering behaviors near EPs, using discrete models of metamaterial (MM) systems. The questions of existence of EPs and their physical manifestations will be addressed with particular focus on symmetry considerations and scattering behavior. Discrete mass-spring models with adjustable parameters are used here to elucidate the EP-related phenomena in a fundamental form. The transfer and scattering matrices are analyzed to provide practical insights on the restrictions associated with reciprocity and fundamental symmetries. By including complex stiffness in frequency domain as a representation of non-conservative mechanical loss or gain, the MM arrays can be tuned to achieve bi-directional transparency or one-way reflection when operating at the EPs. 
This analytical study will contribute to the understandings of EPs in mechanical context and the design of micro-structured media for novel applications.
\PACS{
      {PACS-key}{discribing text of that key}   \and
      {PACS-key}{discribing text of that key}
     } % end of PACS codes
} %end of abstract
\maketitle
\section{Introduction} \label{Intro}

Exceptional points (EPs) were originally introduced~\cite{Heiss1990} in quantum mechanics and are defined as the complex branch point singularities where eigenvectors associated with repeated eigenvalues of a parametric non-Hermitian operator coalesce. This distinguishes an EP from a degeneracy branch point where two or more linearly independent eigenvectors exist with the same eigenvalue. 
The mathematical aspects of EPs have been discussed in literature \cite{Mailybaev2003, Amore2021}.
EPs can widely be found in non-Hermitian systems and have been reported in different physical problems including optics~\cite{Ruter2010,Othman2017} and acoustics~\cite{Lu2018,Maznev2018a,Ding2016}. 
This work aims to use discrete models of mechanical metamaterials (MMs) to analyze the EPs of two different operators (dynamic matrix and scattering matrix) and the associated scattering behaviors. The EPs of the dynamic matrix are shown to lead to bi-directional transparency, which features zero reflection and unitary transmission with zero phase difference. On the other hand, the EPs of the scattering matrix are associated with spontaneously broken parity-time~($\pt$) symmetry and one-way reflection. These two distinct occurrences of EPs have been reviewed and discussed in literature for optical and photonic systems~\cite{Miri2019}. However, there has been little discussion on the EPs in mechanical context.
The introduced discrete systems may be considered as reduced order analogs of continuum micro-structured media and help the conceptual design of these system by removing all but essential dynamic features.

The EPs of a dynamic matrix can be found in the eigenfrequency study, where the equations of motion are established for a repeating unit cell (RUC). Bloch-Floquet condition is embedded in the wavenumber-dependent matrices. Such a setup enables the computation of the eigenfrequencies and mode shapes of an infinite periodic array, for any prescribed wavenumber. The eigenfrequency band structure is of prime importance in the studies of MMs and phononic crystals (PCs)~\cite{Liu2000} as it signifies the overall dispersion of the micro-structured medium. Due to the coupling effects between degrees of freedom in a locally resonant structure, the band structure exhibits mode mixing and frequency band gaps. 
It has been shown~\cite{Amirkhizi2018d} that, an internal resonator does not necessarily lead to a stop band. The existence and the width of a stop band are strongly related to the coupling strength between multiple degrees of freedom. To study this effect with a quantified coupling strength, a tunable discrete model is developed and presented here, in which the coupling tunability is achieved using a skewed resonator. In~\cref{DMEP}, it is shown that a wide band gap is associated with a large coupling constant, while a decoupled resonator leads to independent dispersion branches without a band gap.
In cases where coupling is weak, the band gap becomes extremely narrow and the dispersion curves appear to repel each other to form an avoided crossing. Such a gap could lead to incorrect sorting of branches, due to its extremely small width and the sharp changes in mode shapes. This phenomenon is referred to as level repulsion (LR) and has been studied in literature~\cite{Lu2018, Yeh2016, Wang2016, Amirkhizi2018d}. It is hypothesized that the sharp mode changes may be utilized for accurate and robust identification of a perturbative parameter in the operator under study, in this case wavenumber.

While the frequency dispersion curves (levels) are repelled in the real wavenumber domain, the two dispersion surfaces intersect each other at an EP in the complex domain. Lu and Srivastava~\cite{Lu2018} introduced a method based on the mode shape continuity around such points to distinguish the real vs. avoided crossing points in the band structure. They showed that the instances of frequency level repulsion in the real wavenumber domain have their associated Riemann surfaces crossing at an EP in the complex wavenumber domain. The exotic topology of the eigenvalue surfaces in the vicinity of EPs has attracted extensive research interest in recent years~\cite{Ryu2015, Doppler2016, Xu2016a, Maznev2018a, Shen2018a, Miri2019}. However, the EPs discussed in literature usually possess complex parameters (e.g., frequency, wavevector components) and are studied only in the eigenfrequency analysis. Similarly, in the first part of the present work, complex wavenumber is used as the parameter leading to non-Hermiticity of the dynamic matrix and controls the location of EPs. The question thus rises: how would an EP of the eigenfrequency band structure affect the scattering of such systems? To answer this question, one may seek to tune the parameters that can break the Hermiticity of an elastic system, and then modulate the system so that the complex singularity point is re-positioned onto the real frequency axis. It is shown that wavenumber-parameterized systems can be tuned by adding loss and gain to various spring elements. This could enable moving the location of EP into the real wavenumber domain as well as making the associated frequency to be real (while in contrast the EPs in an earlier work~\cite{Maznev2018a} had complex frequency). Such a system may be studied in a simple harmonic scattering (real frequency) numerical experiment. In practice loss and gain elements may be realized via viscous or other coupled multi-physics (e.g., piezoelectric) components. We derive the conditions to make the EP locate on the real frequency plane and show that stiffness parameters must have certain compatible loss and gain factors. 

With an EP re-positioned to real frequency domain, it is then feasible to analyze the scattering behavior when operating near such an EP. Using the transfer matrix method (TMM), which relates the mechanical states on the left and right boundaries of a finite medium, the scattering coefficients, which describes the relation of incoming and outgoing waves through a sample, can then be derived to study the response when operating near EPs. Transfer matrix method is discussed in depth~\cite{Nemat-Nasser2015, Amirkhizi2017, Nanda2018, Amirkhizi2018c,Psiachos2018a,Psiachos2019} for wave propagation problems in 1D systems. It is widely used to determine the band structure and can also be used to compute the reflectance spectrum~\cite{Ardakani2017}. A similar approach to determine the band gap behavior of permuted PCs is the transfer function method~\cite{AlBabaa2017a,AlBabaa2019}. In~\cref{SecTSM}, the transfer and scattering matrices are constructed for the presented discrete MM array of finite length, with adjustable parameters that allow the unit cell to convert into a monatomic, diatomic, or locally resonant cell. The physical behavior of a MM crystal near an EP (e.g., scattering of steady state waves off a finite specimen) can lead to various interesting phenomena. It is illuminating to summarize the restrictions and simplifications that reciprocity (applicable to all 1D linear systems) and symmetry considerations (applicable to specific structures that admit them) provide, particularly when applied to mechanical systems with loss and gain. 

An example of parity symmetric scattering is shown in~\cref{sec:scatteringexamples}. In this example, the EP of the dynamic matrix is tuned by the loss and gain factors in the viscous or multi-physical springs to have real frequency and wavenumber, so that such a singularity point can be accessed in a scattering experiment.
With certain number of unit cells, the MM sample becomes completely invisible in both directions at the EP frequency of the dynamic matrix, and the energy is dynamically balanced, i.e., the loss and gain mechanisms perfectly cancel each other and total mechanical energy is conserved. The discrete modeling approach helps understand the scattering properties analytically, and can be easily adapted for various tuning possibilities.

On the other hand, if the sample possesses only the combined parity-time ($\pt$) symmetry, then the \emph{scattering matrix} can exhibit EPs as well. To demonstrate this, we show the scattering response of a~$\pt$ symmetric system near the EPs of its scattering matrix spectrum in~\cref{SMEP}. Non-Hermitian Hamiltonians with~$\pt$ symmetry were first discussed by Bender and Boettcher~\cite{Bender1998a}. A more general category of pseudo-Hermitian systems in elastodynamics is investigated by Psiachos and Sigalas~\cite{Psiachos2018a,Psiachos2019}. The asymmetric scattering responses of~$\mathcal{PT}$ symmetric media have been investigated in electronics~\cite{Sakhdari2018}, photonics~\cite{Ge2012}, and acoustics~\cite{Zhu2014, Shi2016, Fleury2016, Achilleos2017, Fleury2014}. These studies have shown that the EPs of scattering matrix correspond to unidirectional zero reflection and unitary transmission. Moreover, the EPs of the scattering matrix are associated with spontaneous symmetry breaking and mark the spectral boundaries between~$\pt$ broken and unbroken phases. The majority of these studies are performed experimentally or numerically using simulations, which can be time consuming or computationally expensive. In contrast, using the analytical formulas derived in~\cref{SecTSM} and appendices, designing these novel artificial media and tuning towards desired target frequencies can be achieved relatively easily.

The feasibility of implementing gain units (represented by complex-valued springs in this work), a necessary ingredient of this study, is a major challenge to experimental realization of such EP-based designs.
To this end, a number of studies have demonstrated implementation of gain units, realized by electronic devices~\cite{Popa2014,Fleury2016} or piezoelectric semiconductors~\cite{Christensen2016}.
In a recent study, Mokhtari \etal~\cite{Mokhtari2020a} show the possibility of accessing EPs with fully elastic PCs in real frequency and wave vector domain. With such a 2D scattering setup, it is then possible to take advantage of the spectral properties of EPs to design novel sensing devices~\cite{wang2021}.

The structure of this paper is as follows. In \cref{DMEP}, we first study the dynamic matrix of a discrete resonator system and show the relationship between coupling strength and level repulsion. This will be followed by an analytical representation and detailed discussion of the eigenfrequency and eigenvector behaviors in the vicinity of EPs. Then it will be shown that by adjusting the stiffness parameters, EPs can be moved to the real frequency and wavenumber domain. In \cref{SecTSM}, the transfer and scattering matrices for discrete MM arrays are presented, along with a discussion on the restrictions reciprocity and fundamental symmetries enforced on these matrices. In \cref{sec:scatteringexamples}, we examine the scattering behaviors of a parity symmetric system and a parity-time symmetric system, which feature bi-directional transparency and one-way reflection near EPs of the dynamic and scattering matrices, respectively. 
The paper is concluded with a summary of important results and potentials for application of EPs resulting from their influence on the physical response of mechanical metamaterials. The use of the discrete mass-spring systems provides fundamental insights in MMs and can be easily adapted for various tuning possibilities.

\section{Exceptional points in the dynamic matrix eigenspectrum}\label{DMEP}
\subsection{Level repulsion and coupling of DOFs} 

\begin{figure}[!ht]
	\begin{subfigure}[b]{0.5\linewidth}
		\centering\includegraphics[height=120pt]{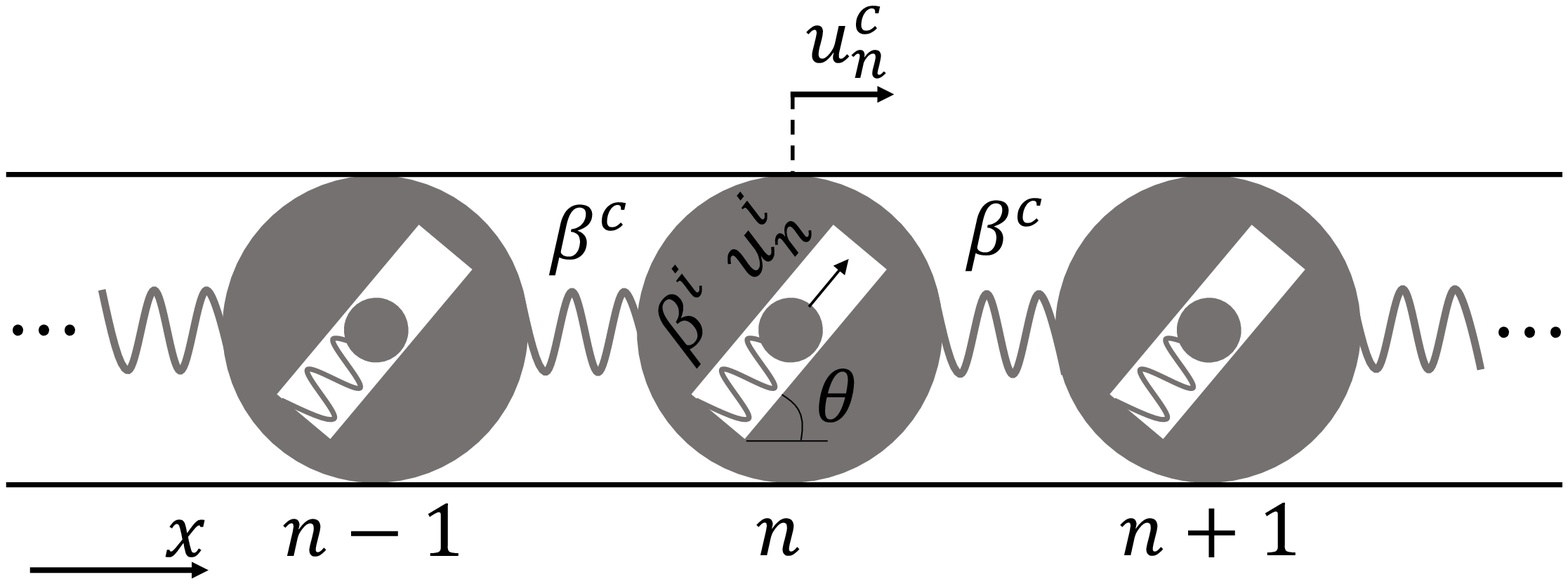}
		\caption{\label{fig:chain}}
	\end{subfigure}%
	\begin{subfigure}[b]{0.5\linewidth}
		\centering\includegraphics[height=150pt]{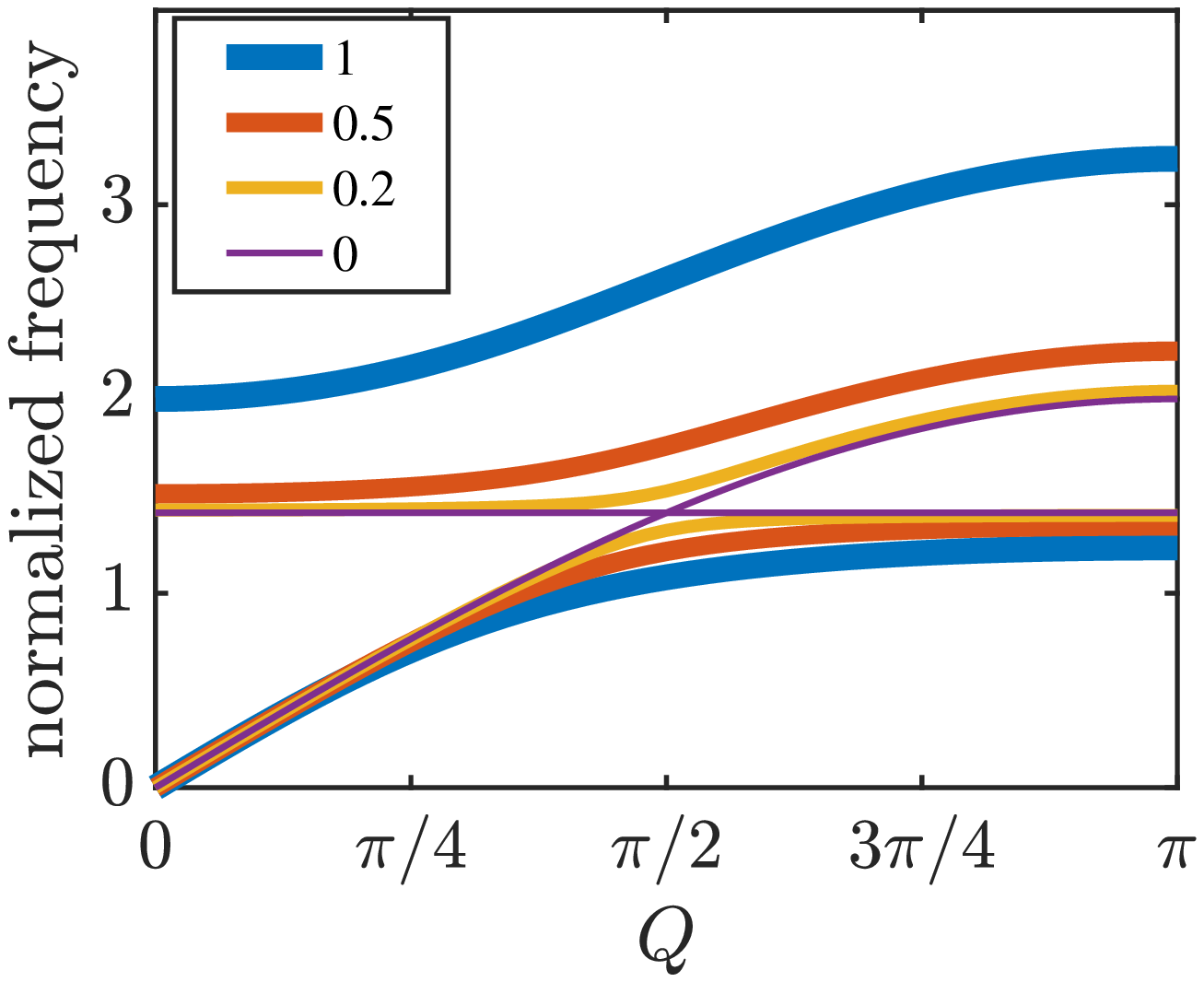}
		\caption{\label{fig:cbandplots}}
	\end{subfigure}
	\caption{(\subref{fig:chain}) Schematic drawing of the studied 1D infinitely periodic resonator array. (\subref{fig:cbandplots})~Longitudinal wave band structure (real domain) for different values of coupling constant~$\kappa=\cos\theta$. In the example here, all parameters (mass, stiffness) are normalized to one. 
	\label{fig:chainlr}} 
\end{figure}

The 1D discrete periodic structure studied in this section is represented in \cref{fig:chain}. Each cell consists of the main ``crystal chain mass''~$M^c$ and the ``internal resonator mass''~$M^i$. In each RUC, a linear spring element with stiffness coefficient~$\beta^i$ connects the resonator to the crystal. There is also a spring with stiffness~$\beta^c$ between every two neighboring crystal masses. To show the level repulsion and EPs with a simple set up, we consider a longitudinal wave propagating along the chain. In this analysis, the crystal masses are constrained to have a single horizontal degree of freedom (DOF). One can assume that the structure is confined in a tube parallel to the~$x$ axis (with frictionless surfaces). It is also assumed that the rotational inertia of the masses are high enough to allow one to ignore the rotational DOFs. The resonator is also constrained to have only one independent DOF,~$u^i_n$, which makes angle~$\theta$ with the horizontal direction and main chain mass DOF,~$u^c_n$. The coupling constant~$\kappa=\cos\theta$ is defined where the angle~$\theta$ is in the range~$[-\pi/2, \pi/2]$. For other values of $\theta$ a simple change of sign in either of the two DOFs will render the following mathematical description identically applicable. When~$\kappa=1$ this model is identical to the 1D lattice with resonator model which can be commonly found in literature \cite{Amirkhizi2018d, Hussein2014a}. 

For the~$n$-th RUC the DOFs that satisfy Bloch-Floquet periodicity can be written as:
\begin{align}\label{eq:uc}u^c_{n}&=u^c\ \exp\left[\mathrm{i}(\omega t-nQ)\right],\\
\label{eq:ui}u^i_{n}&=u^i\ \exp\left[\mathrm{i}(\omega t-nQ)\right],
\end{align}
where~$\omega$ is angular frequency, and~$n$ is an integer representing cell location along the chain. The dimensionless wavenumber~$Q$ represents the phase advance between neighbor cells, and it can be calculated as the product of wavevector component and cell length. In the eigenfrequency study,~$Q$ is usually a prescribed parameter sweeping the Brillouin zone. The complex amplitudes of displacements in harmonic motion~$u^c$ and~$u^i$ are to be determined. To do this, the equations of motion can be written for the~$n$-th cell and resonator (see Appendix~\cref{sec:appEOM}):
\begin{align}\label{eq:1dc}
M^c\frac{\partial^2u^c_n}{\partial t^2}&=\beta^c(u^c_{n+1}-2u^c_n+u^c_{n-1})+\kappa\beta^i(u^i_n-\kappa u^c_n)-(1-\kappa^2) M^i\frac{\partial^2u^c_n}{\partial t^2}, \\
\label{eq:1di}
M^i\frac{\partial^2u^i_n}{\partial t^2}&=\beta^i(\kappa u^c_n-u^i_n),
\end{align}
rendering, for each value of $Q$, an eigenvalue problem:
\begin{align}\label{eq:ruc}
[\vect{D}-\lambda \vect{I}]\vect{U}^R&=\vect{0},\\
\vect{U}^{L\dagger}[\vect{D}-\lambda \vect{I}]&=\vect{0}\label{eq:rucleft},
\end{align}
where~$\lambda=\omega^2$ is the eigenvalue of the dynamic matrix~$\vect{D}=\vect{M}^{-1}\vect{K}$, $\vect{I}$ is the~$2\times2$ identity matrix,~$\vect{U}^R=[u^c, \  u^i]^\top$ is the right eigenvector,~$\vect{U}^L$ is the left eigenvector, and~$\dagger$ denotes complex conjugate transpose.~$\vect{K}$ and~$\vect{M}$ are the stiffness and mass matrices of the cell, respectively: 
\begin{equation}\label{eq:Kmatrix}
\vect{K}=\begin{pmatrix}
4\beta^c \sin^2\dfrac{Q}{2}+\kappa^2\beta^i &-\kappa\beta^i\\
-\kappa\beta^i&\beta^i
\end{pmatrix},
\end{equation}
\begin{equation}\label{eq:Mmatrix}
\vect{M}=\begin{pmatrix}
M^{ci} &0\\
0&M^i
\end{pmatrix}.
\end{equation}
The coupling constant~$\kappa=\cos\theta$ quantifies the interaction strength between the internal resonator and the main crystal chain, and
\begin{equation}
M^{ci}=M^c+(1-\kappa^2)M^i
\end{equation}
is defined as the effective mass associated with ~$u^c_n$ DOF dynamics. Solving the characteristic equation~$|\vect{D}-\omega^2\vect{I}|=0$ yields the frequency band structure, a representation of which is shown in \cref{fig:cbandplots}. In the shown example, all the stiffness and mass values are taken as 1, but the coordinates of a number of important points and other geometrical features can be calculated explicitly in terms of the model parameters. 
As the coupling constant~$\kappa$ approaches 0, level repulsion (LR) becomes more evident and the dispersion curves appear to approach a crossing point. Only when~$\kappa=0$ the resonator and the cell become fully decoupled leading to an actual crossing of the branches. Then a topological transition occurs in the band structure as the frequency gap disappears. 
In such a case, the right eigenvectors on the two crossing dispersion curves will stay linearly independent. 
\newline

\subsection{Exceptional points in the complex band structure}
For a 2-DOF system like the one shown in \cref{fig:chain}, normally there are two eigenfrequencies for each value of wavenumber. There also exist branch points (BP), potentially in the complex domain, where two frequency solutions match (degeneracy or frequency coalescence). After solving for the frequency as a function of~$Q$ analytically, the location of branch points can be obtained:
\begin{align}\label{eq:qep} Q_\mathrm{BP}&=2\arcsin\left(\dfrac{\omega^i}{2\omega^c}\left(1+ \mathrm{i} \kappa\sqrt{\dfrac{M^i}{M^{ci}}}\right)\right),\\
\label{eq:wep}\omega_\mathrm{BP}&=\omega^i\sqrt{1+ \mathrm{i} \kappa\sqrt{\dfrac{M^i}{M^{ci}}}}.
\end{align}
The shown solution is in the region where $\Re Q\geq0$ and~$\Im Q \geq 0$. Here the symbols~$\Re,\Im$ represent the real and imaginary parts of a complex quantity. A non-zero coupling~$\kappa$ will enforce the branch point to be complex-valued for real parameters in the cell and there are in general eight possible solutions, namely~$( \pm Q_{\mathrm{BP}}, \pm \omega_{\mathrm{BP}})$ and $( \pm Q_{\mathrm{BP}}^*, \pm \omega_{\mathrm{BP}}^*)$.
At the branch point \cref{eq:ruc} becomes:
\begin{equation}
\label{eq:dep}
\frac{\kappa \beta^i}{M^{ci}M^i} \begin{pmatrix}
\mathrm{i}\sqrt{M^{ci} M^i} & -M^{i} \\
-M^{ci} &-\mathrm{i}\sqrt{M^{ci}M^i}
\end{pmatrix}
\begin{pmatrix}
u^c\\u^i\end{pmatrix}=\vect{0}.
\end{equation} 
When~$\kappa=0$ the~$2\times2$ matrix in \cref{eq:dep} becomes a zero matrix (which means any arbitrary vector in~$\mathbb{C}^2$ is an eigenvector) and the branch point is exactly the crossing point shown in \cref{fig:cbandplots} residing is in the real domain. The two frequency solutions are overlapping each other while two linearly independent eigenvectors exist. Such a case is referred to as a degeneracy. For non-zero~$\kappa$ values, the branch point is referred to as an exceptional point~\cite{Heiss1990} (EP) which is usually in the complex parameter domain. As the~$\kappa$ value gets closer to zero, avoided crossing/level repulsion will be more apparent in the real domain, and the EP location will have smaller imaginary parts. At the EP there exists only one non-trivial right eigenvector:
\begin{equation}\label{eq:uep}
\vect{U}^R_\mathrm{EP}=\begin{pmatrix}-\mathrm{i} \\  \sqrt{\dfrac{M^{ci}}{M^i}}\end{pmatrix}.\end{equation}
The corresponding left eigenvector of matrix~$\vect{D}$ at EP is 
\begin{equation}\label{eq:ulep}
\vect{U}^{L\dagger}_\mathrm{EP}=\begin{pmatrix} -\mathrm{i}, & \sqrt{\dfrac{M^{i}}{M^{ci}}}\end{pmatrix}.
\end{equation}

\begin{figure}%[!ht]
	\begin{subfigure}[b]{0.5\linewidth}
		\centering\includegraphics[height=159pt]{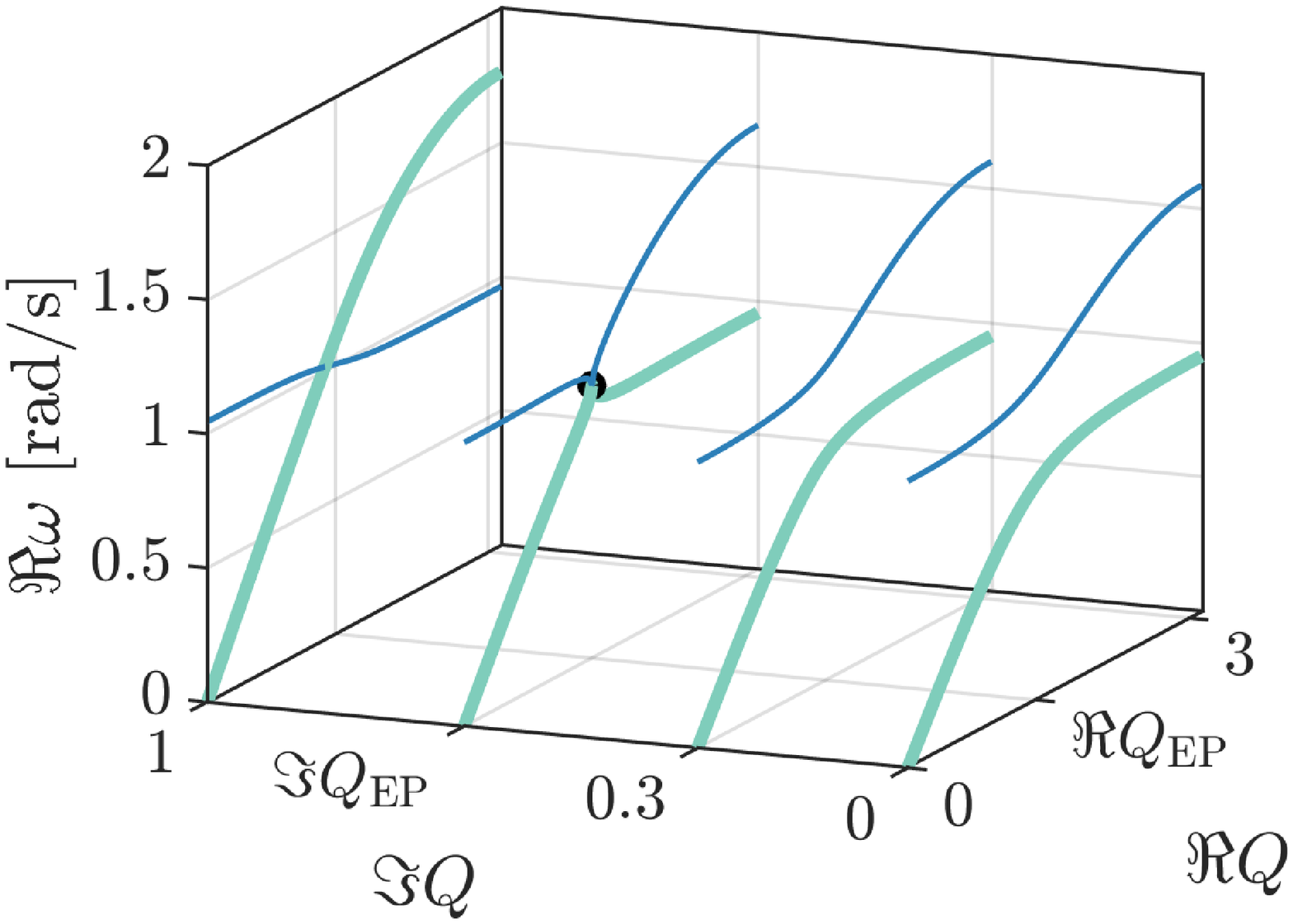}
		\caption{\label{fig:rew}}
	\end{subfigure}%
	\begin{subfigure}[b]{0.5\linewidth}
		\centering\includegraphics[height=159pt]{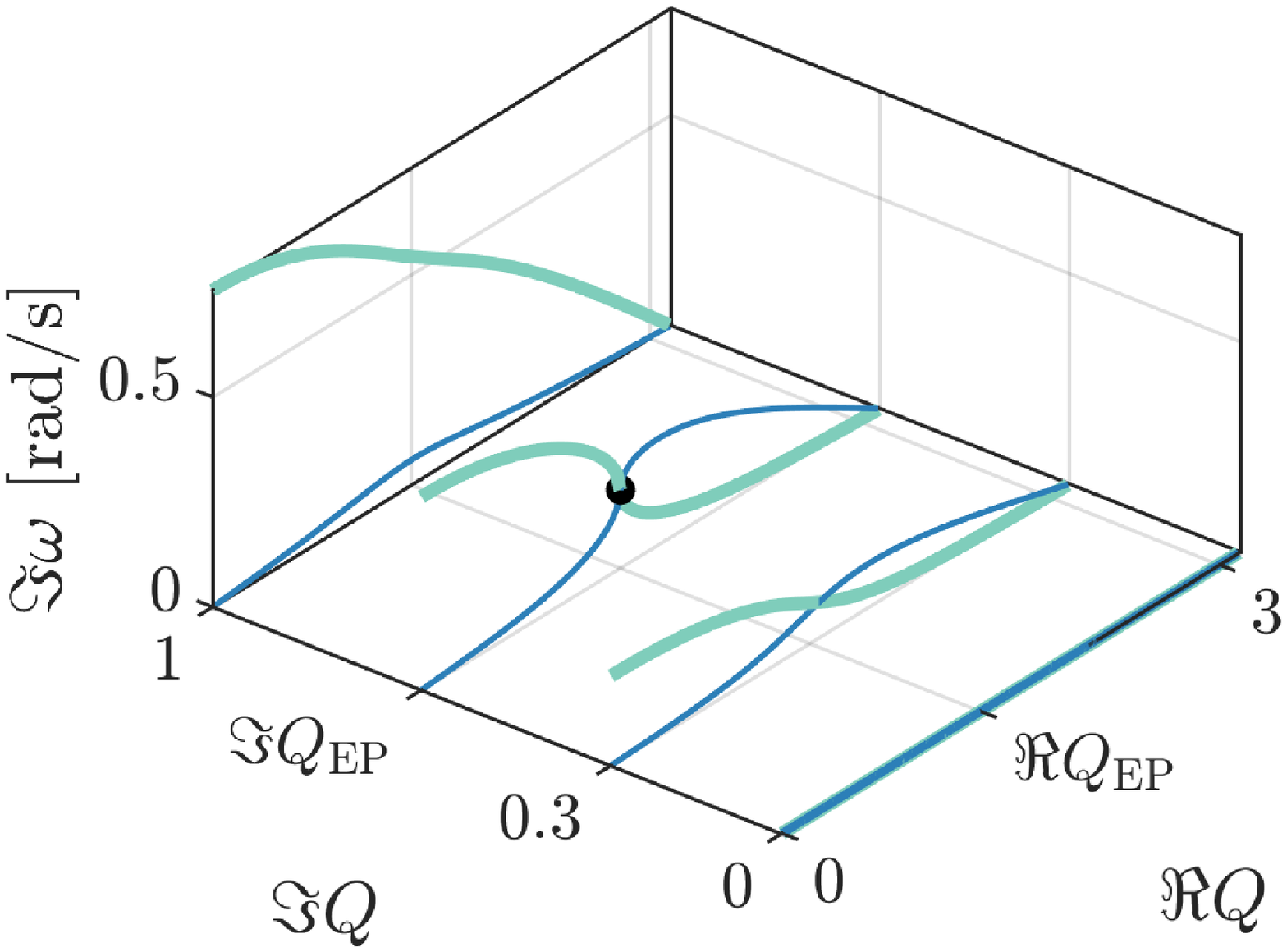}
		\caption{\label{fig:imw}}
	\end{subfigure}
	\begin{subfigure}[b]{0.5\linewidth}
	\centering\includegraphics[height=147pt]{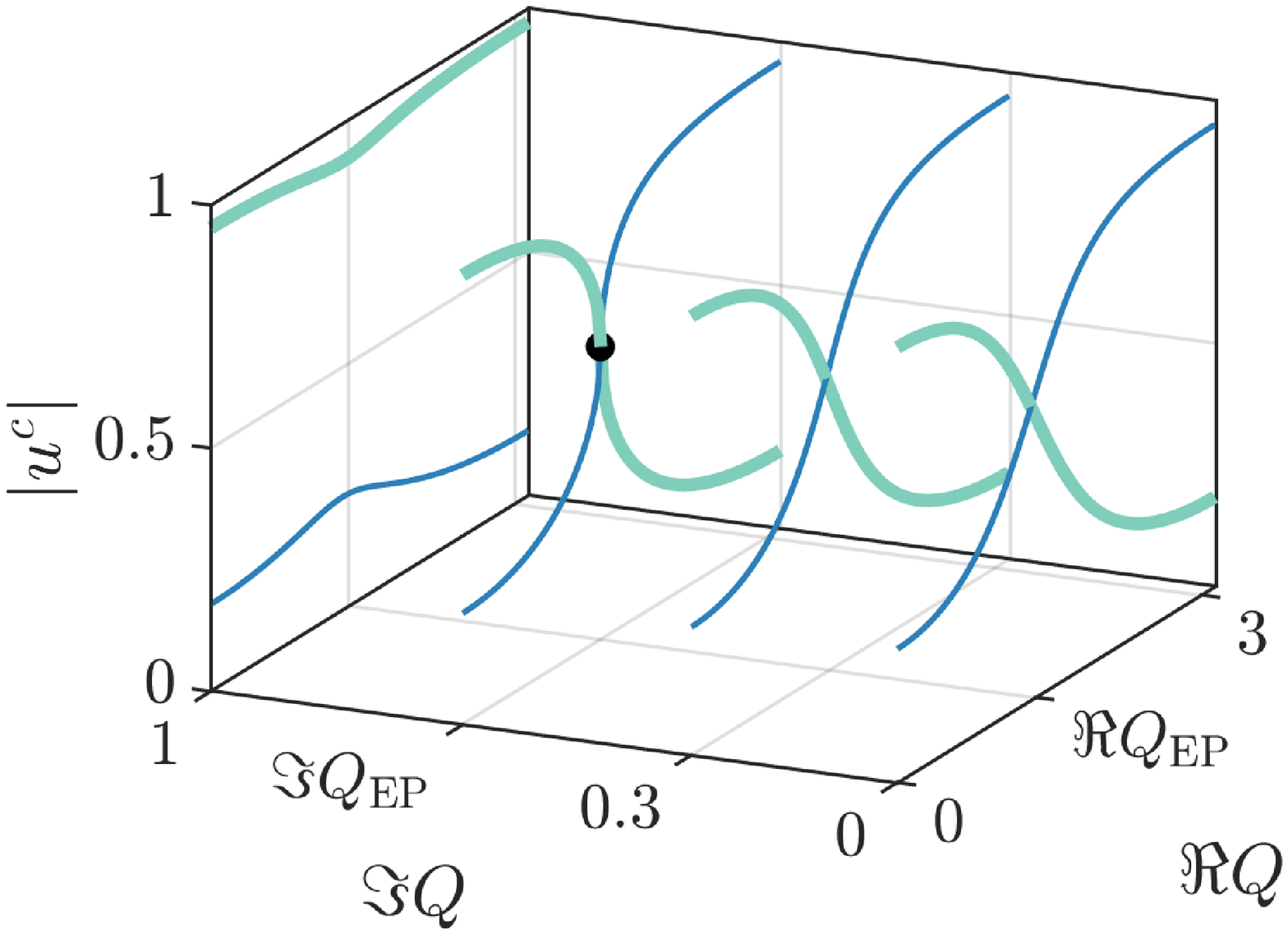}
	\caption{\label{fig:absuc}}
\end{subfigure}%
\begin{subfigure}[b]{0.5\linewidth}
	\centering\includegraphics[height=147pt]{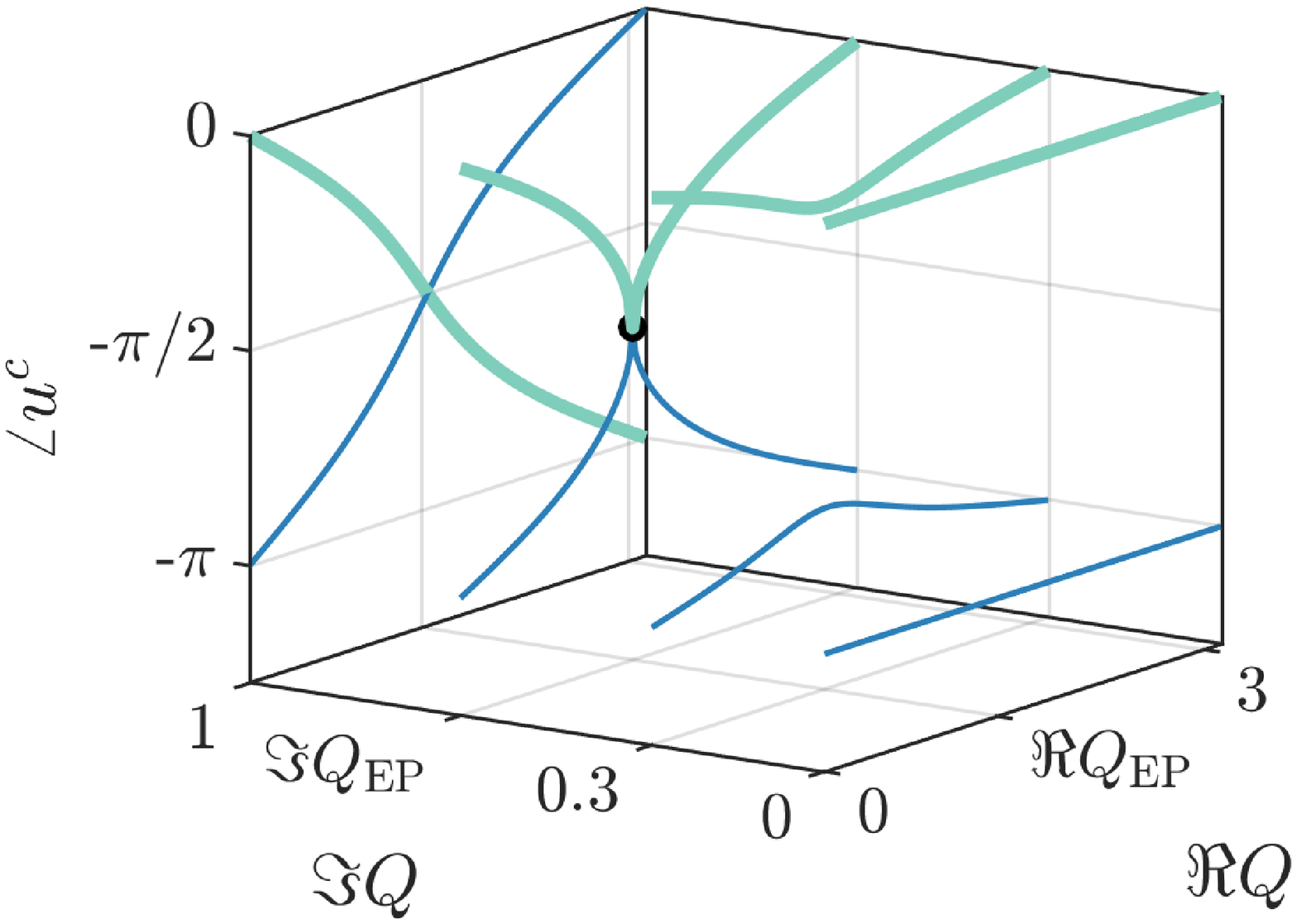}
	\caption{\label{fig:arguc}}
\end{subfigure}
\begin{subfigure}[b]{0.5\linewidth}
	\centering\includegraphics[height=147pt]{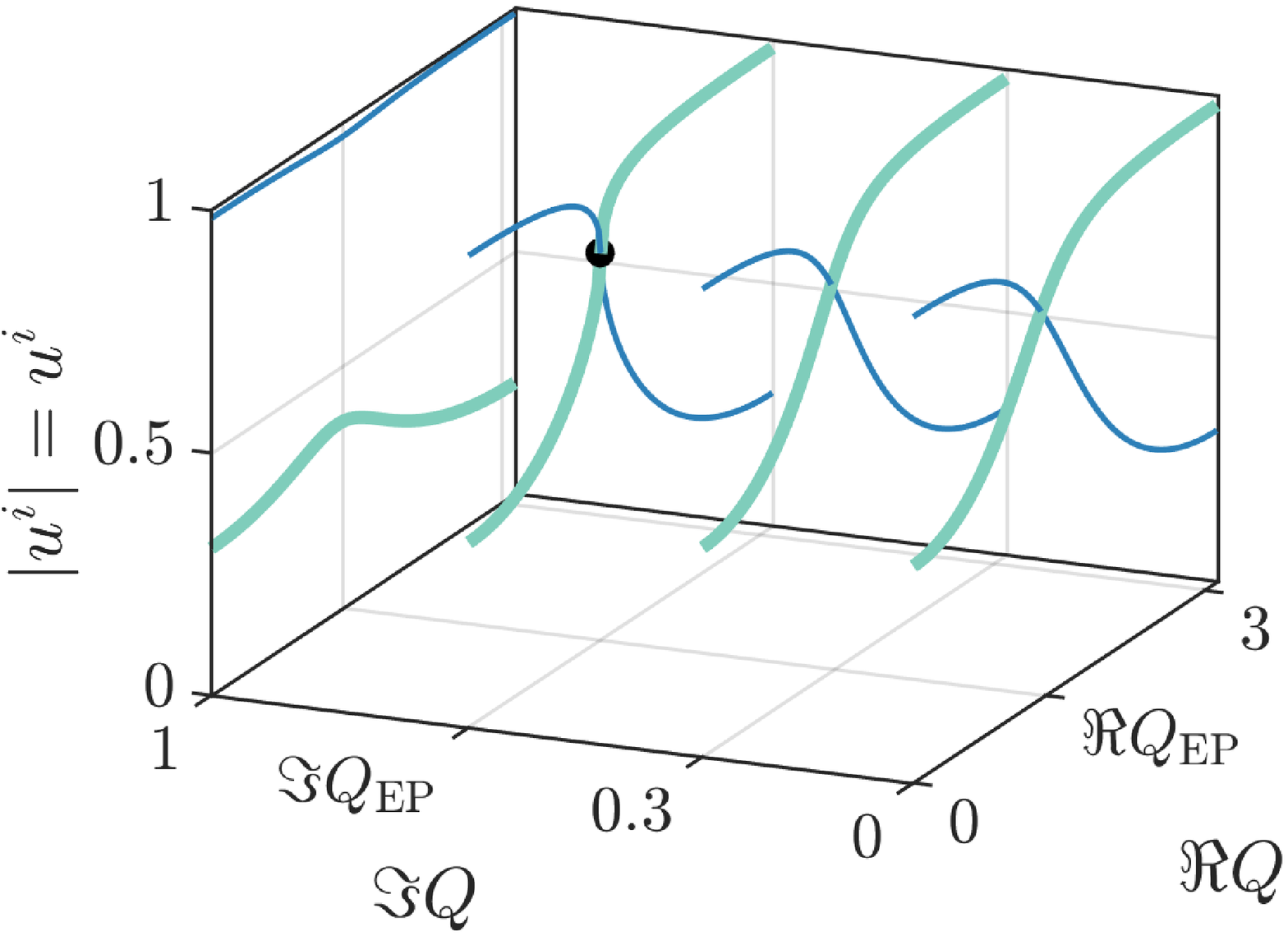}
	\caption{\label{fig:absui}}
\end{subfigure}%
\begin{subfigure}[b]{0.5\linewidth}
	\centering\includegraphics[height=147pt]{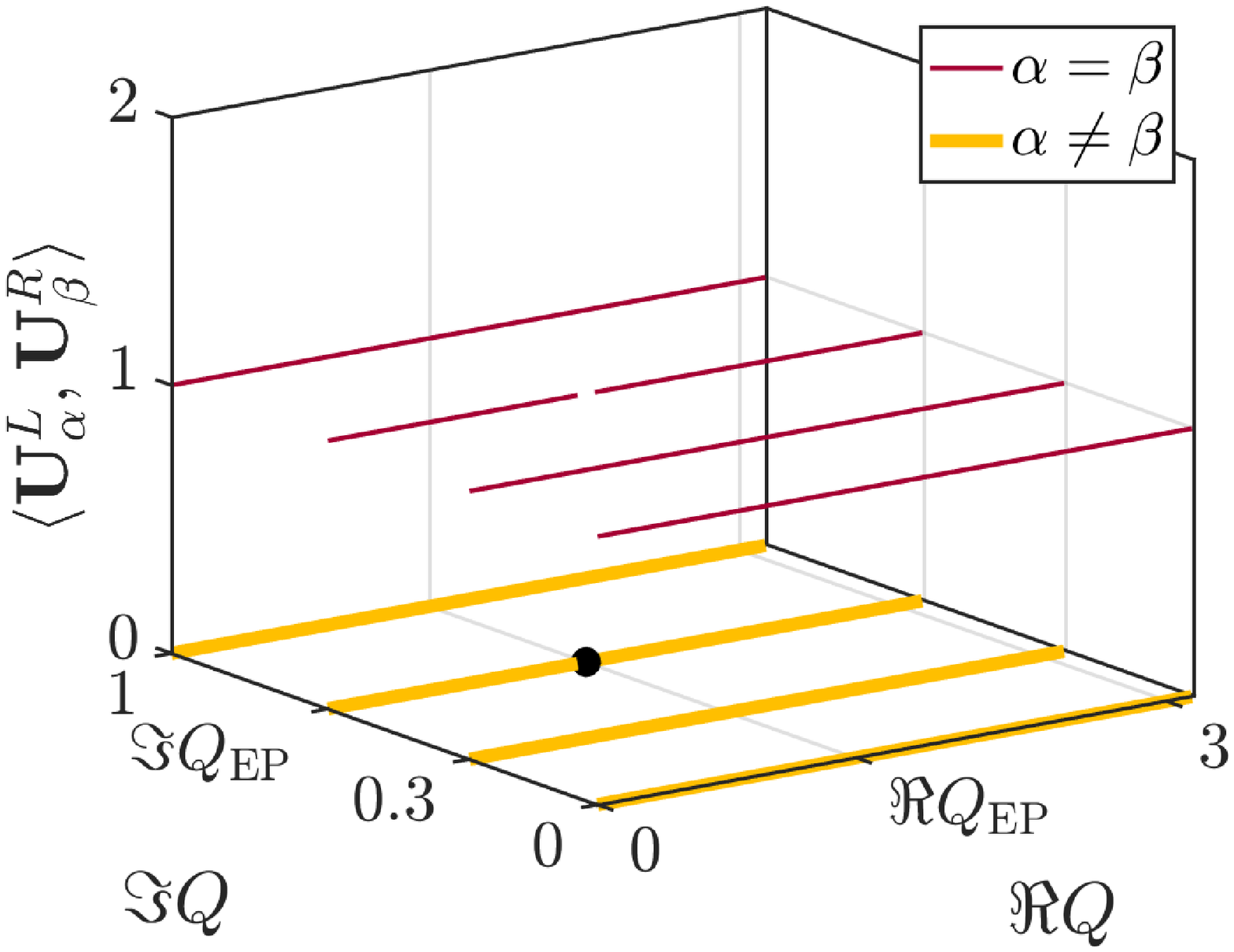}
	\caption{\label{fig:dp}}
\end{subfigure}
	\caption{Four representative cross sectional cuts (at four different~$\Im Q$ values) of the Riemann sheets showing (\subref{fig:rew})~real and (\subref{fig:imw})~imaginary parts of the frequency, (\subref{fig:absuc})~amplitude and (\subref{fig:arguc})~complex argument of main crystal chain DOF~$u^c$, and (\subref{fig:absui})~resonator DOF~$u^i$ (when eigenvectors are normalized for it to be real). The two modes are sorted based on branch continuity along $\Re Q$. (\subref{fig:dp}) Inner product of left and right eigenvectors which belong to same ($\alpha=\beta$) and different ($\alpha\neq \beta$) modes. \label{fig:cband}} 
\end{figure}

Here we show an example of the complex band structure for~$\kappa=0.5$ and allow the wavenumber~$Q$ to be complex. All cell parameters (mass, stiffness) are set to one for the sake of demonstration. The calculated complex frequency and right eigenvectors are shown only in region where~$\Re Q\geq0, \Im Q \geq 0$, and $\Re \omega \geq 0$. \Cref{fig:rew,fig:imw} show the real and imaginary part of the band structure, respectively. The components of the right eigenvector are shown in~\cref{fig:absuc,fig:absui}. For this configuration the EP is located at~$Q_\mathrm{EP}=1.36218+\mathrm{i}0.63297$ and~$\omega_\mathrm{EP}=1.01711+\mathrm{i}0.18580$, as represented by the solid black dot. The two modes are separated based on the continuity of branches in any complex $Q$ disk around the origin that does not include $Q_{\mathrm{EP}}$ and are shown in different colors. The corresponding right eigenvector components are shown in \cref{fig:absuc,fig:absui}. The complex eigenvectors associated with each mode are normalized by a complex factor in such a way that
$\norm{\vect{U}^R}=\sqrt{u^{c*} u^c+u^{i*} u^i}=1$, and~$u^i\in \mathbb{R}$. It is important to normalize both the amplitude and complex phase of the eigenvector in such a way to ensure consistency throughout the analysis. To keep~$u^i$ on the real axis, both components of eigenvector are rotated together in the complex plane keeping their ratio unchanged. The major benefit of phase normalization is that the eigenvector components can be shown in a continuous manner (see \crefrange{fig:absuc}{fig:absui}) even in the vicinity of the EPs, thus making it easier to understand the mode shape behavior. 
It is clear that both the frequencies and eigenvectors form Riemann sheet structures in the vicinity of the EP, and all the complex quantities can be made to behave continuously (when properly normalized) with respect to~$\Re Q$ and~$\Im Q$, at any simply connected neighborhood that does not include $Q_\mathrm{EP}$. For the sake of presentation quality of 3D figures, we show only four representative cuts of these Riemann sheets at~$\Im  Q=0,\ 0.3,\ \Im Q_\mathrm{EP},$ and 1. 
When~$\Im Q=0$ the frequencies are real. The two branches are clearly distinguished by considering $\Re \omega$ and $\angle u^c$. As~$\Re Q$ increases, the amplitudes of eigenvector components increase or decrease monotonically due to local resonance. There is an inverse correlation between the coupling strength (which is quantified as~$\kappa$ in this case) and the abruptness of such change in amplitude. Weaker coupling (smaller but non-zero~$\kappa$) results in more evident frequency level repulsion and sharper changes in displacement amplitudes. The exact~$\pi$ difference between the two lines in \cref{fig:arguc} indicates a sign difference between the~$u^c$ of acoustic and optical branches, with~$u^i$ normalized to be real and positive. A detailed discussion on eigenfrequency and mode shape behaviors in the LR region in real~$Q$ domain can be found in our previous work~\cite{Amirkhizi2018d}. 

As~$\Im Q$ increases from 0, all the presented quantities show similar trends in terms of continuity. The two branches remain continuous, as long as~$\Im Q<\Im Q_\mathrm{EP}$.
When~$\Im Q=\Im Q_\mathrm{EP}$, all these quantities coalesce at the EP. Continuity with respect to~$\Re Q$ can not be used as the basis of branch selection at this point due to the coalescence. In other words, branch sorting becomes ambiguous at~$\Im Q=\Im Q_\mathrm{EP}$. If one seeks to extend the continuous branches (for $\Im Q < \Im Q_\mathrm{EP}$) beyond $\Im Q_\mathrm{EP}$, by maintaining continuity along $\Im Q$, the resulting choices will be discontinuous along $\Re Q$ when $\Im Q > \Im Q_\mathrm{EP}$. If one wishes to maintain the continuity along $\Re Q$, the branches will be discontinuous along $\Im Q$, when $\Re Q > \Re Q_\mathrm{EP}$, see for example the slice at~$\Im Q=1$ in \crefrange{fig:rew}{fig:absui}. In general, it can be seen that both the eigenvalues and the eigenvectors maintain analyticity, except at the EP where the Taylor series expansion fails.

The calculated inner products of the normalized left and right eigenvectors are shown in \cref{fig:dp}, where it can be seen that the left and right eigenvectors corresponding to the different eigenfrequencies are orthogonal to each other, except at the EP. The bi-orthogonality relation of the non-Hermitian system reads:
\begin{equation}
\langle\vect{U}^L_\alpha,\vect{U}^R_\beta\rangle=\begin{cases} 
0 & \text{at EP;}\\
\delta_{\alpha\beta} & \text{elsewhere;}
\end{cases}
\end{equation}
where~$\delta_{\alpha\beta}$ is the Kronecker delta, and subscripts~$\alpha,\beta=1,2$ denote the first or second mode. The self-orthogonality~\cite{NimrodMo} at the EP implies a defect of the Hilbert space~\cite{Kato,Rotter2003}.

\subsection{Tuning an EP into real frequency and wavenumber domain}
All previous results are based on an EP with complex~$Q$ and~$\omega$ values. It is possible to look for EPs with real~$Q$ using complex stiffness~$\beta$. For physical realization of such systems, see \cref{Intro}. Associated Riemann sheets for the band structure and eigenvectors may be found following the procedure discussed earlier. A similar analysis on bifurcation in the vicinity of EPs in such systems with damping (complex stiffness) can be found in Ref.\cite{Maznev2018a}, where the EP has complex frequency and real wavenumber. If one wishes to study the scattering properties around an EP, one may try instead to design a system including an EP that has both real frequency and $Q$. Such a system may be interrogated experimentally through scattering of harmonic waves off a finite specimen. Given the EP location in \cref{eq:qep} and \cref{eq:wep}, it is possible to make both~$Q_\mathrm{EP}$ and~$\omega_\mathrm{EP}$ real if the complex arguments of stiffness parameters are such that:
\begin{equation}
%\angle \beta^i&=\angle \left[\left(1\pm \mathrm{i}c \sqrt{\frac{M^i}{M^{ci}}}\right)^{-1}\right],\label{epbi}\\
\angle \beta^c=-\angle \beta^i=\angle \left[1+ \mathrm{i}\kappa \sqrt{\frac{M^i}{M^{ci}}}\right].\label{eq:epbc}
\end{equation}
It is inevitable that such process will make one stiffness parameter lossy while the other requires gain (i.e., has negative imaginary part), which is feasible as discussed in literature\cite{Popa2014,Fleury2016,Christensen2016}.
With these adjustments in stiffness values, \crefrange{eq:qep}{eq:ulep} are still valid, and the eigenfrequency and eigenvector behaviors are still qualitatively the same as those shown in \cref{fig:cband}, except the location of EP is now adjustable. The behavior of such structures will be further studied in \cref{sec:scatteringexamples}.

The measurable response of the micro-structured media is, in fact, not just based on their dispersion surfaces, but rather more thoroughly understood from the scattering off finite specimens. In the following sections, the influence of EPs of the band structure, and independently, those of the scattering matrix, on the overall response of finite specimens are studied. 
\newline

\section{Transfer and scattering matrices}\label{SecTSM}

\begin{figure}[!h]
	\centering\includegraphics[width=240pt]{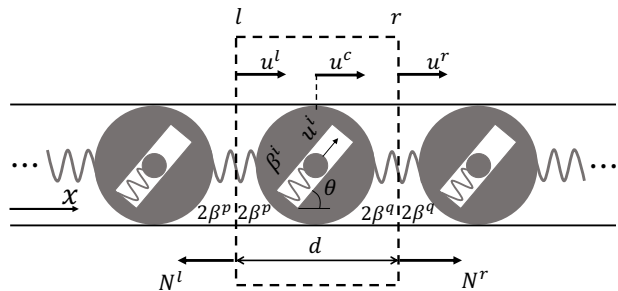}
	\caption{\label{fig:TMRUC} Control volume selection for transfer matrix analysis. Note that the cell is no longer inherently symmetric. }
\end{figure}

The analysis of steady state waves (real frequency) traveling in an infinite homogeneous domain and interacting with a finite sized specimen of a 1D MM array (with finite number of unit cells) can be solved easily using the transfer matrix (TM) of such structures. This is different from eigenfrequency analysis which analyzes an infinite periodic array of unit cells, though the eigenfrequency band structure are also essentially associated with the eigenvalues of the transfer matrix. In either case, the TM is the matrix form of the linear relationship between the physical states on the left and right boundaries of a control volume or the unit cell. To utilize the transfer matrix method, the unit cell is selected as the part in the dashed rectangle shown in \cref{fig:TMRUC}. It is selected in such a way that the springs connecting crystals are cut in the middle. To be more general, the springs at the left and the right sides of a crystal chain atoms are allowed to be different, as denoted by stiffnesses~$\beta^p$ and~$\beta^q$ in \cref{fig:TMRUC}.
Therefore, the left and right halves of the main crystal chain springs are $2\beta_p$ and $2\beta_q$. This setup allows the analysis of non-periodic/permuted MM samples. The derivation of the TM can be found in Appendix~\cref{appTM}. Applying Bloch-Floquet periodicity in an infinitely periodic array, the governing equation reads:
\begin{equation}
\label{eq:cellTMeig}
[\TM^{cell}-\mathrm{e}^{-\mathrm{i}Q}\vect{I}]
\begin{pmatrix}
v^l\\N^l\end{pmatrix}=\vect{0},
\end{equation} 
where~$\TM^{cell}$ is the transfer matrix of a unit cell,~$Q$ is the phase advance, and~$v^l,N^l$ are the velocity and internal axial force at the left boundary.
For any desired frequency, solving the characteristic equation of \cref{eq:cellTMeig} for the normalized wavenumber $Q$ yields the band structure. This is complementary to eigenfrequency calculation, where one would solve for frequencies given a prescribed normalized wavenumber (phase advance), $Q$. 

\begin{figure}[!h]
	\centering\includegraphics[width=300pt]{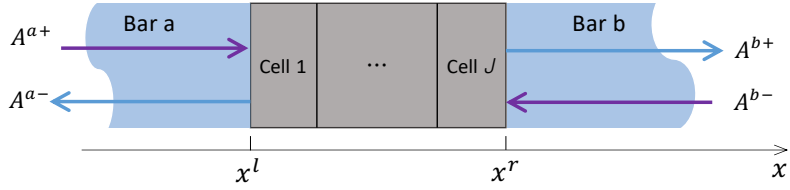}
	\caption{\label{fig:bar} Scattering set up of~$J$ cells between two semi-infinite domains. 
	}
\end{figure} 

For finite structures, once the TM of one arbitrary cell or a number of them is obtained, one can retrieve the scattering coefficients of the model as shown in \cref{fig:bar}. In the scattering experiment, the sample is set between two homogeneous semi-infinite domains and contains~$J$ cells. The cells of the sample are in general allowed to be different. Here the semi-infinite domains, without loss of generality, are modeled as circular bars. The Young's modulus, mass density, and cross sectional radius of the two identical cylindrical bars are denoted by~$E_0$,~$\rho_0$, and~$r_0$, respectively. We generally equate the measurement/de-embedding locations with the boundaries of the sample, i.e., $x^a = x^l$ and $x^b = x^r$. Note that the superscripts~$l$ and~$r$ here represent locations with respect to the entire sample and their distinction from the cell faces earlier should be clear from the context. The displacements at these locations are assumed to have the form~$A^{(a,b)(+,-)} e^{\ii\omega t}$ (harmonic waves), where the superscripts~$+$ and~$-$ represent the waves propagating in positive and negative~$x$ directions in the bars (in the sense of phase advance or flux direction). The scattering matrix 
\begin{equation}\label{eq:sm}
\TSM= 
\begin{pmatrix}
\SM_{ba} & \SM_{bb}  \\
\SM_{aa} & \SM_{ab}
\end{pmatrix}
\end{equation}
relates the complex displacement or velocity amplitudes of outgoing ($A^{b+}$ and~$A^{a-}$) and incoming ($A^{a+}$ and~$A^{b-}$) waves at measurement locations~$x^l$ and~$x^r$:
\begin{equation}\label{eq:asa}
\vect{A}_{out}=\begin{pmatrix}
A^{b+}\\A^{a-}\end{pmatrix}=
\TSM
\begin{pmatrix}
A^{a+}\\A^{b-}
\end{pmatrix}=\TSM\vect{A}_{in}.
\end{equation}
The derivation of the scattering matrix can be found in Appendix~\cref{appSM}.

If one considers~$\abs{\TM}=1$ and~$Z^a=Z^b$, then~\crefrange{eq:saa}{eq:dels} will be simplified and identical to the scattering coefficients derived in literature~\cite{Amirkhizi2017}, and the scattering matrix $\TSM$ given by \cref{eq:sm} has the eigenvalues:
\begin{equation}\label{eq:seiga}
\sigma_{1,2}=\SM_{ab}\pm\sqrt{\SM_{aa}\SM_{bb}},
\end{equation}
since~$\SM_{ab}=\SM_{ba}$. 
The eigenvectors are
\begin{equation}\label{eq:seige}
\vect{\mu}_{1,2}=\begin{pmatrix}\pm\sqrt{\SM_{bb}}\\\sqrt{\SM_{aa}}\end{pmatrix}.
\end{equation}
Coalescence of the $\TSM$ eigenspectrum occurs if and only if one of the reflection coefficients becomes zero. An EP of the scattering matrix is thus related to one-way reflection phenomenon. 

\subsection{Effect of reciprocity}\label{sec:recipeffects}

Reciprocity is commonly considered~\cite{Cebrecos2019a, Horsley2014, Fleury2015a, Deak2012} as a property observed in scattering measurements, i.e., the equivalence of transmission coefficients when a source and a detector exchange their positions. 
It is shown in Ref~\cite{Alizadeh2021} that elastodynamic reciprocity imposes certain restrictions on the constitutive tensors of layered media. 
Here, we consider reciprocity as a fundamental property of linear elastic structures (and a broad class of linear viscoelastic systems, e.g., those that can be represented by simple networks of 2-force spring or dashpot, or continuum elements with isotropic viscoelasticity), described by Betti's reciprocity theorem.
It can be seen from~\cref{eq:TMRUC} that this transfer matrix has a determinant of 1, i.e.,~$\abs{\TM^{cell}} = 1$. The unimodularity of the TM is equivalent to the reciprocity of the 1D medium. It is worth to study the proof and understand the assumptions that are sometimes implicitly included. Consider an arbitrary control volume in a 1D linear time-invariant system. Assuming absence of body forces, Betti's reciprocity theorem in frequency domain states that:
\begin{equation}\label{eq:betti}
	F^l_\alpha u^l_\beta + F^r_\alpha u^r_\beta = F^l_\beta u^l_\alpha + F^r_\beta u^r_\alpha,
\end{equation}
where superscript $l$ or $r$ represent the left or right boundary of the control volume and subscripts $\alpha$ and $\beta$ represent two different states, and $u$ and $F$ represent displacement and applied force on the boundaries. The axial traction force (positive in the positive $x$ direction) are $F^l=-N^l$ and $F^r=N^r$, and $N^{l,r}$ are assumed to be tensile positive. The theorem is generally proved in elastic systems based on the existence of the strain energy density function and the subsequent major symmetry of the elasticity tensor~\cite{Achenbach2006}. In linear viscoelastic systems, isotropic material response will also allow for a proof in frequency domain. Analytical considerations have been used to address this in viscoelastic materials~\cite{Rogers1963,Day1971,Matarazzo2001}.
For the systems studied here (among a large class of discrete structures), the reciprocity can be proven explicitly. The general proof is omitted, but we show that the reciprocity of the systems considered here is equivalent to the $\abs{\TM} = 1$. Assuming harmonic velocities $v^{l,r}=\ii \omega u^{l,r}$,~\cref{eq:betti} becomes:
\begin{equation}\label{eq:reci}
	v^r_\alpha N^r_\beta - v^r_\beta N^r_\alpha = v^l_\alpha N^l_\beta - v^l_\beta N^l_\alpha~.
\end{equation}
Define matrix $\vect{J}$ as 
\begin{equation}\vect{J}=\begin{pmatrix}
	0 &1\\ -1&0
	\end{pmatrix}~,
\end{equation}
and
\begin{equation}\vect{\psi}=\begin{pmatrix}
	v\\ N
	\end{pmatrix}~.
\end{equation}
Then the reciprocity condition \cref{eq:reci} can be written as
\begin{equation}\label{eq:reci2}
	(\TM \vect{\psi}^l_\alpha)^\top \vect{J} (\TM \vect{\psi}^l_\beta)=(\vect{\psi}^l_\alpha)^\top \vect{J} \vect{\psi}^l_\beta~.
	\end{equation}
Since the L.H.S. of \cref{eq:reci2} is identical to $(\vect{\psi}^l_\alpha)^\top (\TM^\top \vect{J}\TM) \vect{\psi}^l_\beta$, the structure is reciprocal if and only if TM is symplectic, i.e.,
\begin{equation}\label{eq:symp}
	\TM^\top\vect{J}\TM=\vect{J},
\end{equation}
which is equivalent to $\abs{\TM}=1$ for $2\times2$ matrices.

The transfer matrix of any array of cells may be constructed by multiplying their individual transfer matrix in reverse order since, $\vect{\psi}^r$ of one cell is the same as $\vect{\psi}^l$ of the cell to its right. Therefore, the transfer matrix of any array of such cells, will also be reciprocal and have a determinant of 1. This would be true regardless of whether either the parity symmetry (i.e., $\beta^p=\beta^q$) or time-reversal symmetry (i.e., $\{\beta^p,\ \beta^q\}\subset\mathbb{R}$) of any of the cells is broken or not. Since $\abs{\TM} = 1$, the two eigenvalues of~$\TM$ are inverse of each other, which yields~$Q^-(\omega) = -Q^+(\omega)$, where the superscripts~$+$ and~$-$ represent up to two possible solutions, presumably forward and backward traveling waves in the sense of phase advance or power flux. Note that at least for 1D media, either $\mathcal{P}$ or $\mathcal{T}$ symmetry will inherently lead to reciprocity, but a system that only admits combined $\mathcal{PT}$ symmetry can potentially be non-reciprocal.

\subsection{Symmetry considerations}\label{sec:symmetryeffects}
The effect of symmetries of the sample are best represented in the transfer matrix formulation and, similar to the reciprocity consideration, they are independent of the environment. However, the effect on scattering response will include the environment properties. 
The summary of symmetry restrictions is listed in~\cref{table:1}. Here, 
$$
\vect{F}=\begin{pmatrix}
-1 & 0\\
0&1
\end{pmatrix},~
\vect{P}=\begin{pmatrix}
0 & 1\\
1 & 0
\end{pmatrix}.
$$
The detailed proof is omitted, but can be derived following similar processes presented in literature~\cite{Jin2016, Ge2012, Mostafazadeh2014}.
Coalescence of the $\TSM$ eigenvectors and its inherent one-way reflection phenomenon can not be achieved by a parity symmetric system (including the identical bars on either side) because~$\mathcal{P}$ symmetry simply imposes strong condition that enforces equal reflections, i.e.,~$\SM_{aa}=\SM_{bb}$. A system that has time reversal~$\mathcal{T}$ symmetry does not support the coalescence of $\TSM$ eigenvectors either because it leads to $|\SM_{aa}|=|\SM_{bb}|$, which again precludes one-way reflection. However, a~$\pt$ symmetric system that lacks individual $\mathcal{T}$ and $\mathcal{P}$ symmetries may demonstrate non-trivial coalescence of~$\TSM$ eigenvectors and the one-way reflection phenomenon. While~$\pt$ symmetry is usually studied with electromagnetic/optical setups~\cite{Hu2017b,Suneera2018}, the conclusions here apply as well.

\begin{table}[h!]
\centering
\caption{Effects of symmetries on the transfer and scattering matrices. 
Notice that the~$\TM$ restrictions of the~$\mathcal{P}$ symmetry implicitly leads to reciprocity through~$|\TM|=1$ as well, which is why it is not shown in the table, but in case of ~$\mathcal{T}$ it is an additional conclusion.}
\label{table:1}
%\begin{ruledtabular}
%\begin{tabular}{lcr}
\begin{tabular}{lll} %
\hline\noalign{\smallskip} %
Symmetry      & $\TM$ restrictions     & $\TSM$ restrictions  \\ 
%\hline
\noalign{\smallskip}\hline\noalign{\smallskip}
$\mathcal{P}$  & $\TM=\vect{F}\TM^{-1}\vect{F}$ & $\TSM=\vect{P} \TSM \vect{P} $ \\ 
$\mathcal{T}$  &   $ \TM=\vect{F}\TM^{*}\vect{F}$, $|\TM|=1$                 &      $\TSM=\vect{P} (\TSM^{*})^{-1} \vect{P}$               \\ 
$\mathcal{PT}$ &   $ \TM=(\TM^*)^{-1} $                  &     $\TSM=(\TSM^*)^{-1}$                \\
%\end{tabular}
%\end{ruledtabular}
\noalign{\smallskip}\hline
\end{tabular}
\end{table}

The summary here is applicable for 1D linear media. The excitation frequencies are prescribed to be real. Although this paper mainly uses discrete systems as examples, the conclusions here are independent from the discrete setup and are applicable for 1D continuum systems as well. One can use these relations to simplify the process and reduce the number of needed experiments in real scattering experiments~\cite{Aghighi2019,Abedi2020}. In the following, we choose~$Z^a=Z^b=Z_0=\pi r_0^2 \sqrt{E_0 \rho_0}$ so that we can focus on the symmetry properties of the MM samples only.
\newline

\section{Examples of scattering responses}\label{sec:scatteringexamples}
In the following illuminating examples the same setup in \cref{SecTSM} is used, where the bars have Young's modulus~$E_0=\SI{69}{GPa}$, density~$\rho_0=\SI{2710}{kg/m^3}$, and radius~$r_0=\SI{1}{mm}$. The unit cell length is~$d=\SI{0.1}{m}$. Different cases are demonstrated by varying the number and properties of the unit cells.

\subsection{Scattering at the EP of dynamic matrix spectrum ($\mathcal{P}$ symmetric system)
}
\label{EGEP}

\begin{figure}[!h]
	\begin{subfigure}[b]{0.4385\linewidth}
		\centering\includegraphics[height=140pt]{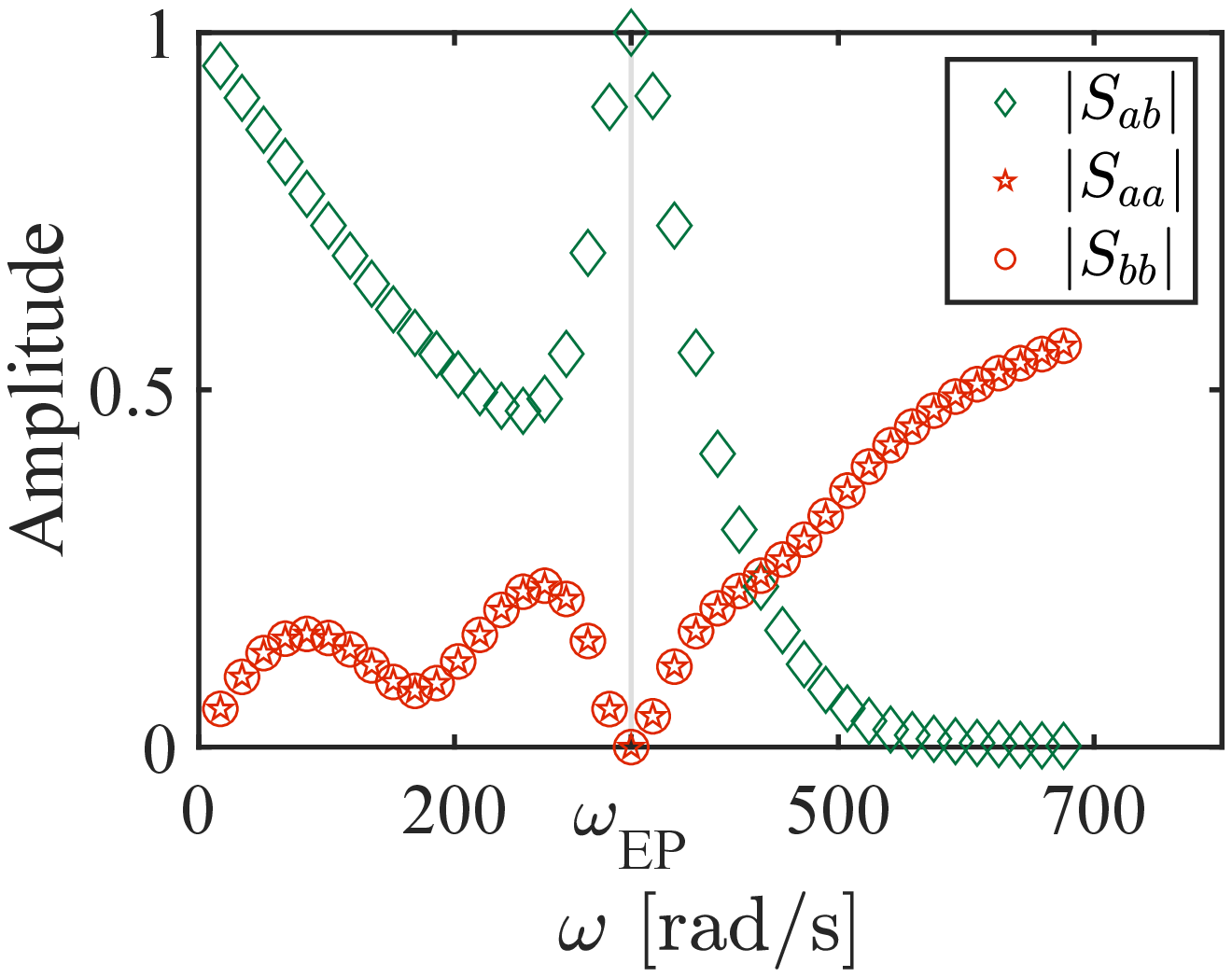}
		\caption{\label{fig:d4am}}
	\end{subfigure}%
	\begin{subfigure}[b]{0.4385\linewidth}
		\centering\includegraphics[height=140pt]{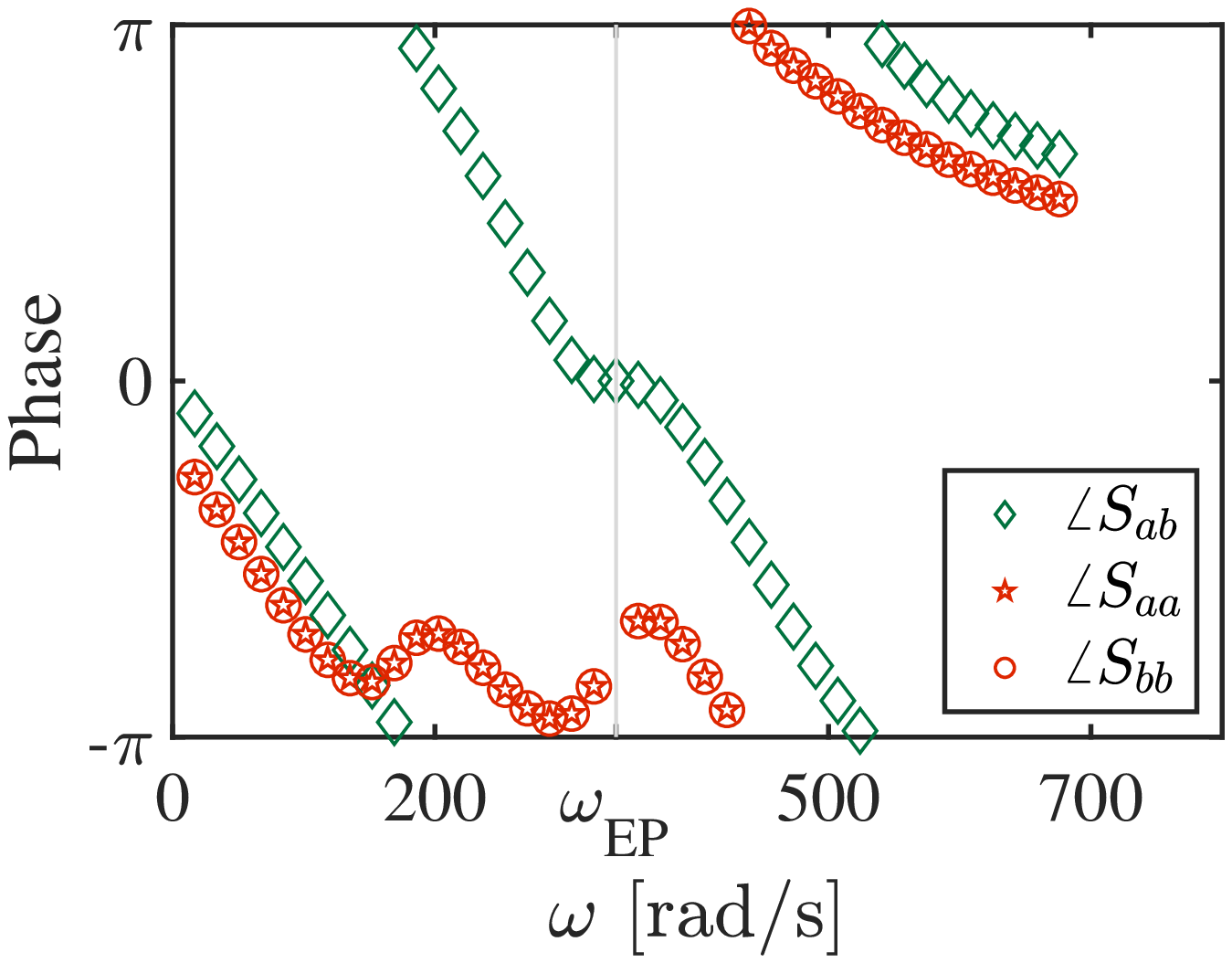}
		\caption{\label{fig:d4ph}}
	\end{subfigure}
	\begin{subfigure}[b]{0.4385\linewidth}
		\centering\includegraphics[height=140pt]{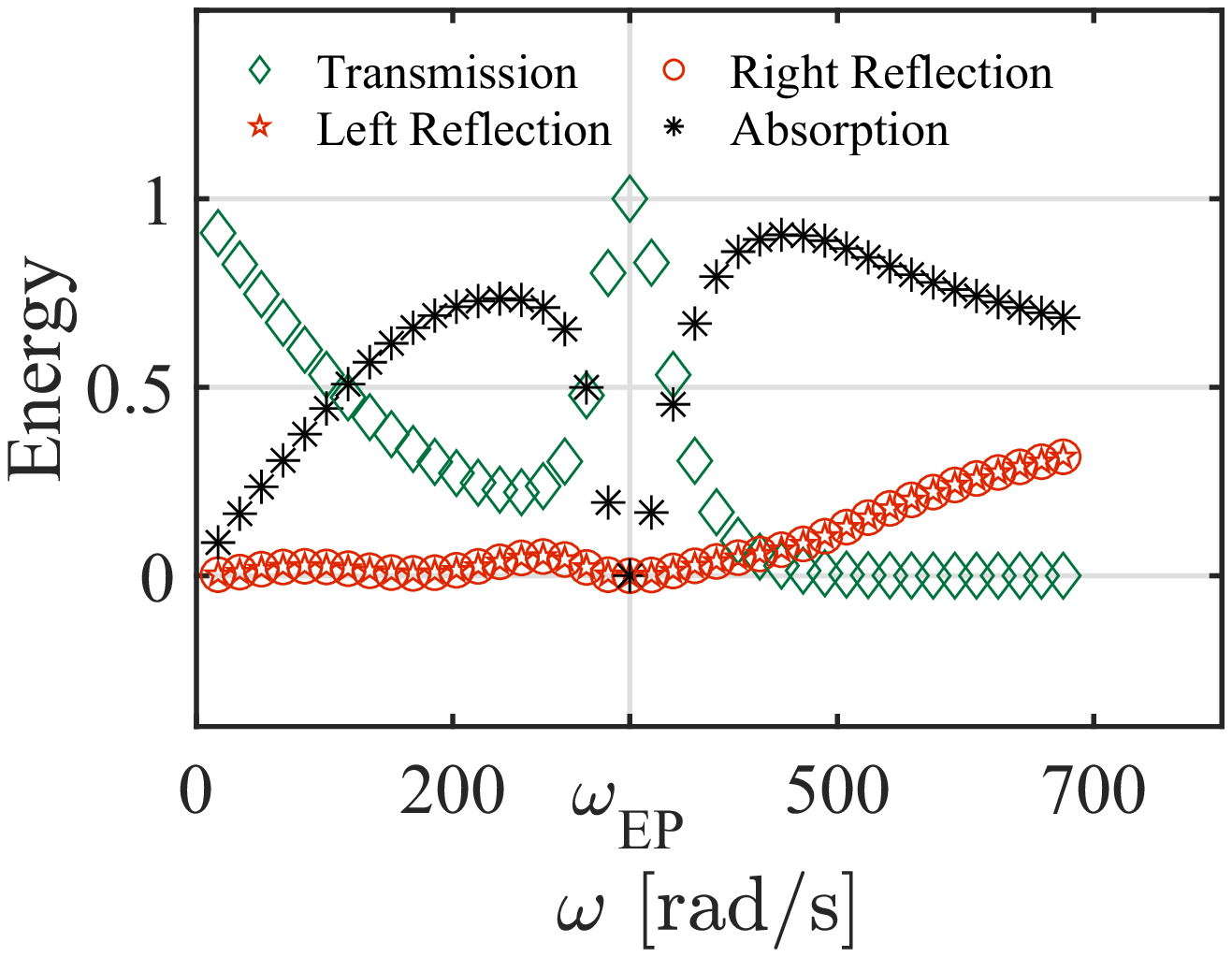}
		\caption{\label{fig:d4en}}
	\end{subfigure}%
	\begin{subfigure}[b]{0.4385\linewidth}
		\centering\includegraphics[height=140pt]{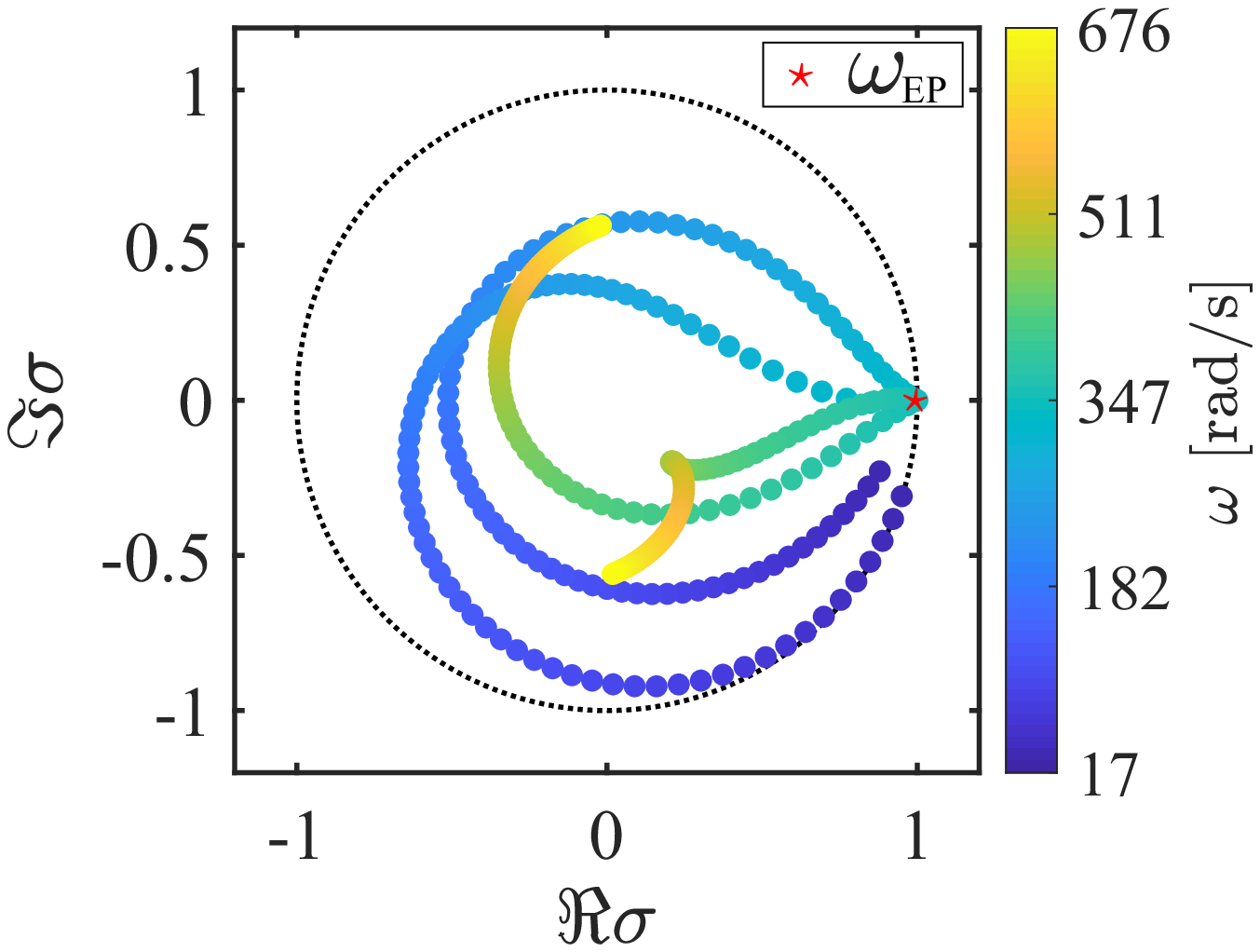}
		\caption{\label{fig:d4sig}}
	\end{subfigure}
	\begin{subfigure}[b]{0.89\linewidth}
		\centering\includegraphics[height=140pt]{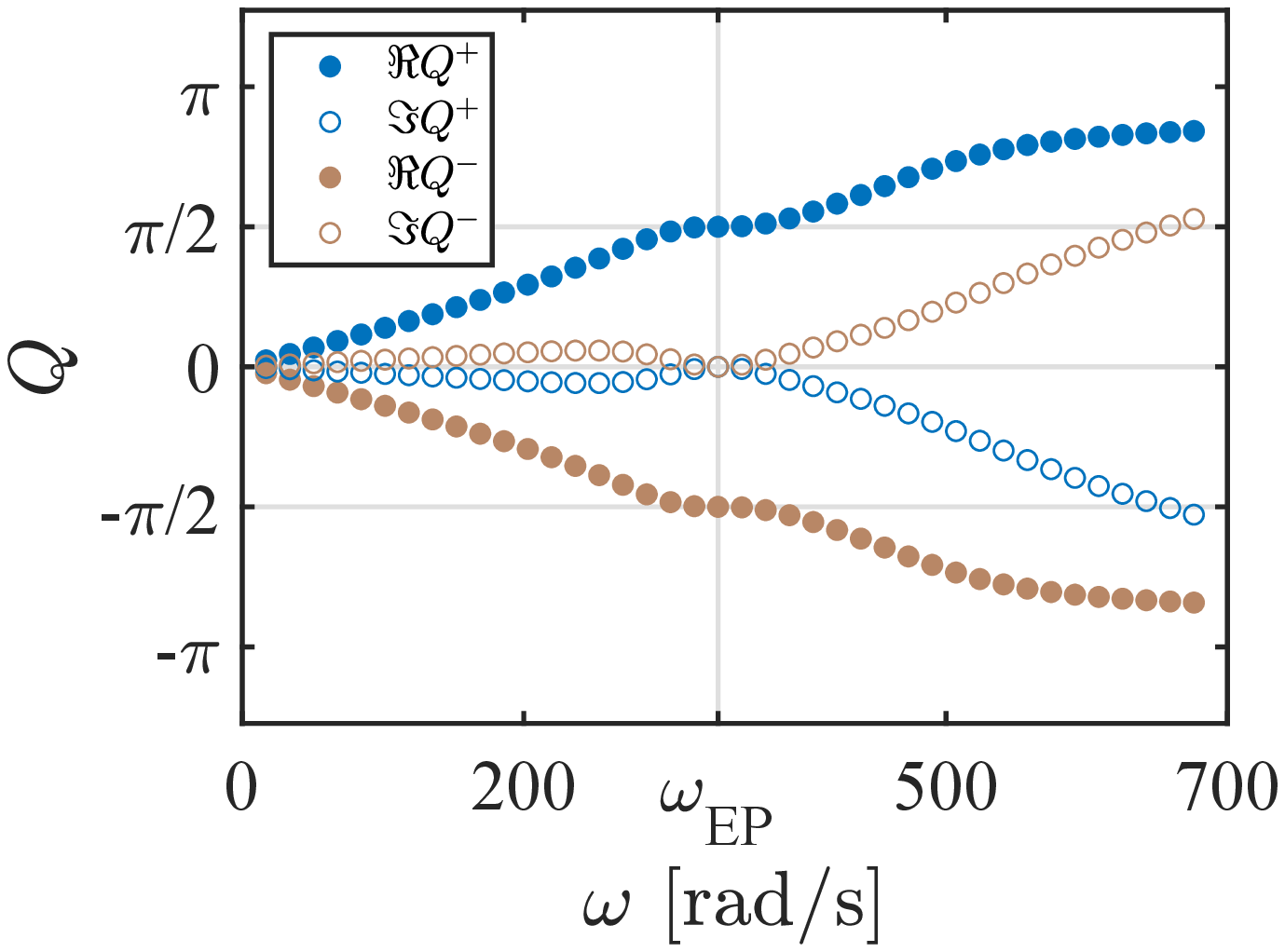}
		\caption{\label{fig:d4bs}}%
	\end{subfigure}
	\caption{Scattering response of the sample with 4 cells. The exceptional point belongs to the eigenfrequency band structure and is moved to real frequency domain ($\omega_\mathrm{EP}\approx\SI{338.1}{rad/s}$) by using suitable loss and gain springs. The band structure shown in~(\subref{fig:d4bs}) is calculated using TMM. The scattering amplitudes, phases, and power fluxes (as well as net power loss) are shown in ~(\subref{fig:d4am}),~(\subref{fig:d4ph}) and~(\subref{fig:d4en}), respectively. The eigenvalues of~$\TSM$, denoted by $\sigma$, are shown in~(\subref{fig:d4sig}). 
	\label{fig:D4EP}} 
\end{figure}

In this section, we examine the scattering properties at the EP of the dynamic matrix eigenspectrum derived in~\cref{DMEP}. To study the scattering of steady state harmonic waves, the EP is located in the real domain. 
In this section the unit cell is set to be symmetric (i.e.,~$\beta^p=\beta^q=\beta^c$, see \cref{fig:TMRUC} and \cref{eq:TMRUC}). The main chain crystal and resonator masses are chosen to be equal~$M^c=M^i=\SI{0.1}{kg}$ and the coupling constant is selected as~$\kappa=0.5$. The stiffness values are~$\beta^p=\beta^q=\beta^{i*}\approx\SI{10+i3.78}{kN/m}$, based on \cref{eq:epbc}. The band structure of such a unit cell can be obtained using \cref{eq:cellTMeig}, and is shown in \cref{fig:d4bs}. The two solutions are denoted by~$Q^\pm$. The EP is calculated based on \crefrange{eq:qep}{eq:wep} and is located at~$\omega_\mathrm{EP}\approx \SI{338.1}{rad/s}$, and~$Q_\mathrm{EP}^{\pm}=\pm\pi/2$, which are both real due to the choice of $\beta$ values.
As pointed out by Maznev~\cite{Maznev2018a}, no branch bifurcation can be observed due to the fact that the wavenumber~$Q$ is solved as a complex function of real frequency in the scattering analysis. 
This band structure is associated with an infinitely periodic array and therefore agnostic to the number of cells in the scattering analysis. However, the actual scattering responses of such structures are affected by the number of unit cells.

To demonstrate this a sample consisting of~$J=4$ unit cells is analyzed. The amplitude, phase, and associated power flux scattering coefficients are shown in \crefrange{fig:d4am}{fig:d4en}. The eigenvalues~$\sigma$ of the scattering matrix are shown in \cref{fig:d4sig}. At the frequency of the EP, unitary transmission and zero reflection can be observed. The energy is dissipated at most frequencies due to the lossy nature of cell springs~$\beta^p$ and~$\beta^q$. The net absorption (loss) of energy per unit time is shown by in \cref{fig:d4en}. At the EP frequency, the power generated by the gain in resonator springs compensates exactly for the loss from main chain springs. This fact can also be seen in the eigenvalue plot \cref{fig:d4sig} as only at the EP frequency, the two eigenvalues of~$\TSM$ reach the unit circle (shown as the dashed circle), indicating the amplitude of incoming and outgoing waves are equal.

In the cases studied here, the phases of the stiffness parameters follow the relation in \cref{eq:epbc}, while ~$\Re \beta^i=\Re \beta^c$ and~$M^i=M^c$. Then based on \cref{eq:qep}, \cref{eq:wep} and \cref{eq:TMRUC} it can be found that at the EP the sample transfer matrix becomes
\begin{equation}
\TM(\omega_\mathrm{EP})=(-\vect{I})^{N/2},\label{eq:TSEN}
\end{equation}
when~$N$ is even. 
Under such circumstances, the transmission coefficients~$\SM_{ab}$ and~$\SM_{ba}$ will have unitary amplitudes, and the reflection coefficients~$\SM_{aa}$ and~$\SM_{bb}$ will vanish. This bi-directional reflectionlessness at~$\omega_\mathrm{EP}$ is not affected by the outside material (assuming two bars are identical), as all the impedance terms in the S-parameters will be canceled out. The frequency-dependent reflectionless scattering property can lead to design of wave filtering devices that only allow waves with certain frequency to transmit. According to \cref{eq:TSEN}, the sample becomes fully invisible at~$\omega_\mathrm{EP}$ if cell number is~$J = 4, 8, 12, \cdots$, as if the two boundaries~$x^l$ and~$x^r$ are directly connected, irrespective of the outside bar material. 
However, for~$J=2,6,10\cdots$ cells, an extra phase of~$\pi$ will be added to each scattering coefficient, with the amplitude being the same as observed also in~$J = 4, 8, 12, \cdots$ cases.

At the frequency of the dynamic matrix EP, the sample with even $N$ happens to satisfy an apparent overall~$\mathcal{T}$ symmetry condition for $\TSM$, with balanced energy gain and loss in the system. Multiple application scenarios could arise with such exotic properties of EPs. For example, the bi-directional reflectionless features of EPs can be used for acoustic camouflage. Since the reflected waves will be suppressed at the EPs’ frequencies, a target object covered by a properly designed micro-structured medium will be undetectable by a sonar-based sensor. 
The scattering matrix has repeating eigenvalues at this frequency, but the two eigenvectors remain linearly independent and, not surprisingly, there is no reason for the scattering matrix to have an EP related to that of the dynamic matrix. 
The EP and coalescence of the scattering matrix eigenspectrum shall be analyzed next.
 
\subsection{Scattering at the EP of scattering matrix spectrum ($\pt$ symmetric system)}\label{SMEP}

The scattering matrix may have an EP only when the system exhibits $\pt$ symmetry but not individual $\mathcal{P}$ or $\mathcal{T}$ symmetries. Since the EP of the scattering matrix is generally unrelated to the eigenfrequency band structure of locally resonant systems and to simplify further exposition, we remove the internal resonator so that~$M^i = 0$. Consider two spring constants~$\beta^q=\beta^{p*}$ with $\beta^p$ in the first quadrant (lossy). A cell with~$\beta^q$ on right and~$\beta^p$ on the left and mass $M^g$ is referred to as~$g$. The transfer matrix of the~$g$ cell is:
\begin{equation}\label{eq:TMRUCba}
\TM^{(g)}=
\begin{pmatrix}
1-\dfrac{M^{g}\omega^2}{2\beta^q} & \mathrm{i}\omega\dfrac{2\beta^p+2\beta^q-M^{g}\omega^2}{4\beta^p\beta^q}  \\
\mathrm{i}M^{g}\omega & 1-\dfrac{M^{g}\omega^2}{2\beta^p}
\end{pmatrix}~.
\end{equation}
A cell with~$\beta^p$ on the right and~$\beta^q$ on the left and mass $M^l$ is referred to as~$l$. The transfer matrix of the~$l$ cell is denoted as~$\TM^{(l)}$, which can be derived by changing the superscript~$p$ into~$q$ and vice versa in \cref{eq:TMRUCba} as well as changing $M^g$ to $M^l$. We construct a sample consisting of five cells, for which total transfer matrix is~$\TM=\TM^{(g)}\TM^{(l)}\TM^{(g)}\TM^{(l)}\TM^{(g)}$. A numerical example based on ~$\beta^p = (10+1\ii) \ \mathrm{kN/m}$ and~$\beta^q = (10-1\ii) \ \mathrm{kN/m}$ with masses~$M^{g} = \SI{0.12}{kg}$ and~$M^{l} = \SI{0.10}{kg}$ is studied here.

\begin{figure}[!h]
	\begin{subfigure}[b]{0.4385\linewidth}
		\centering\includegraphics[height=140pt]{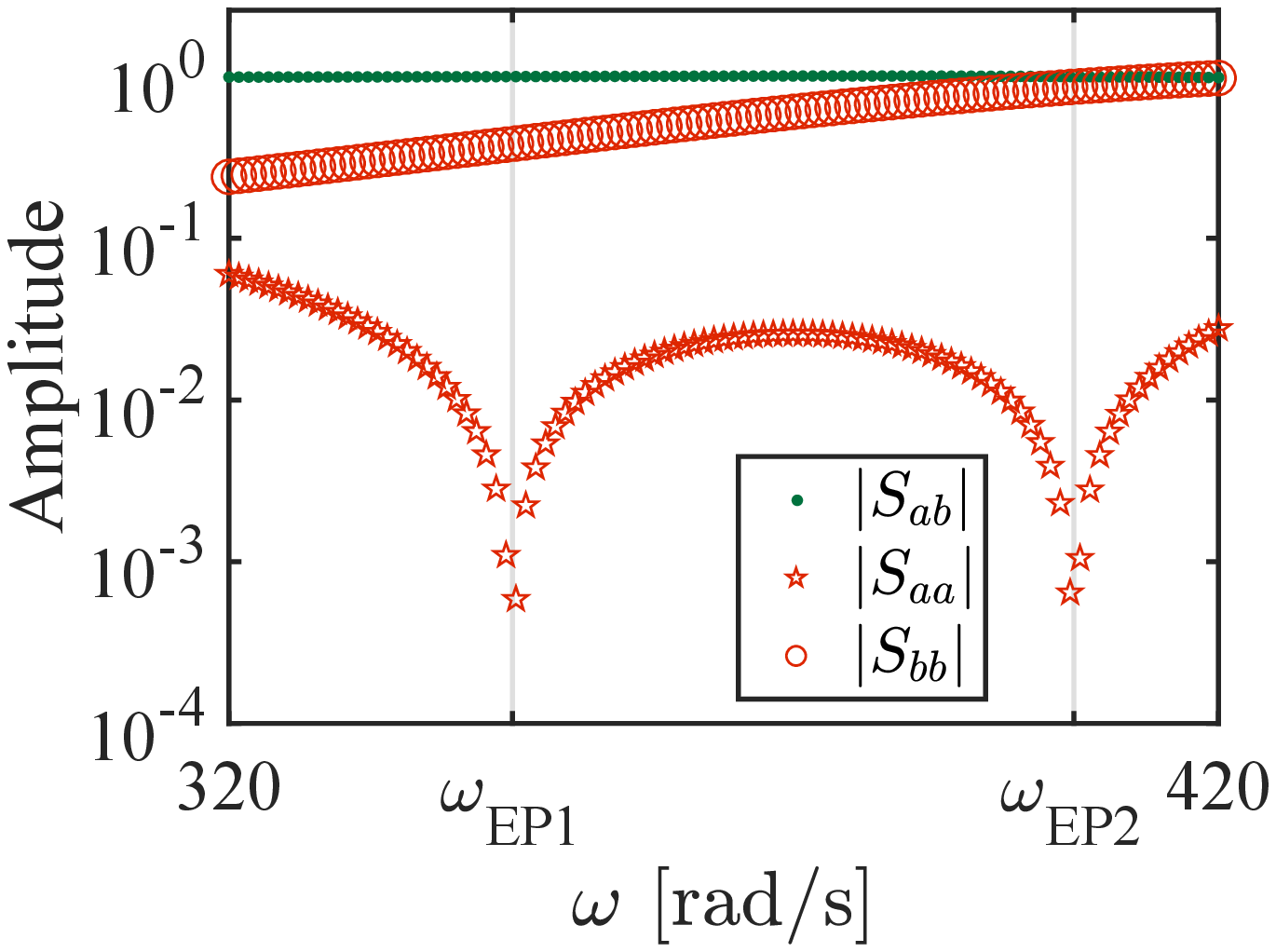}
		\caption{\label{fig:slam}}
	\end{subfigure}%
	\begin{subfigure}[b]{0.4385\linewidth}
	\centering\includegraphics[height=140pt]{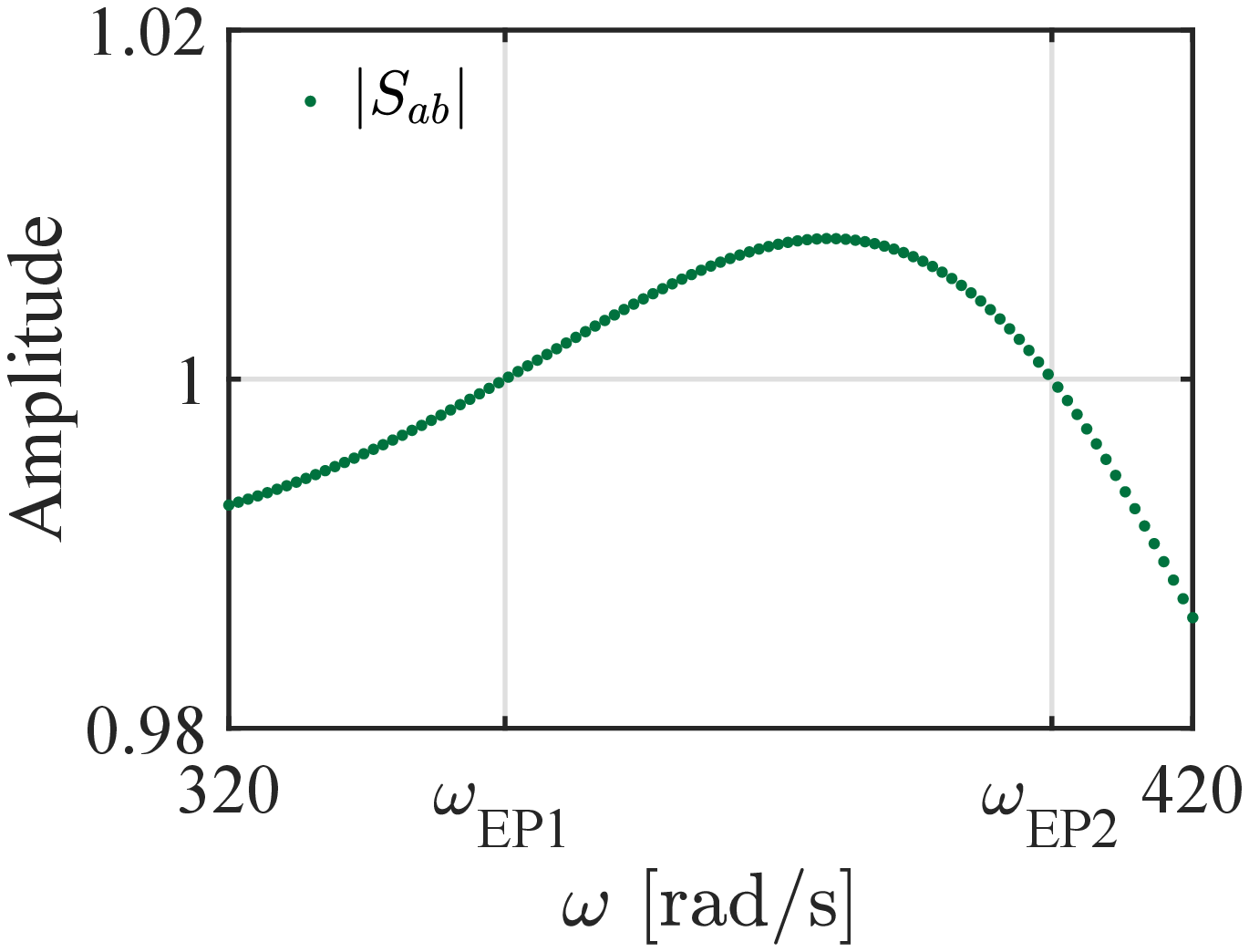}
	\caption{\label{fig:stram}}
    \end{subfigure}
	\begin{subfigure}[b]{0.4385\linewidth}
		\centering\includegraphics[height=140pt]{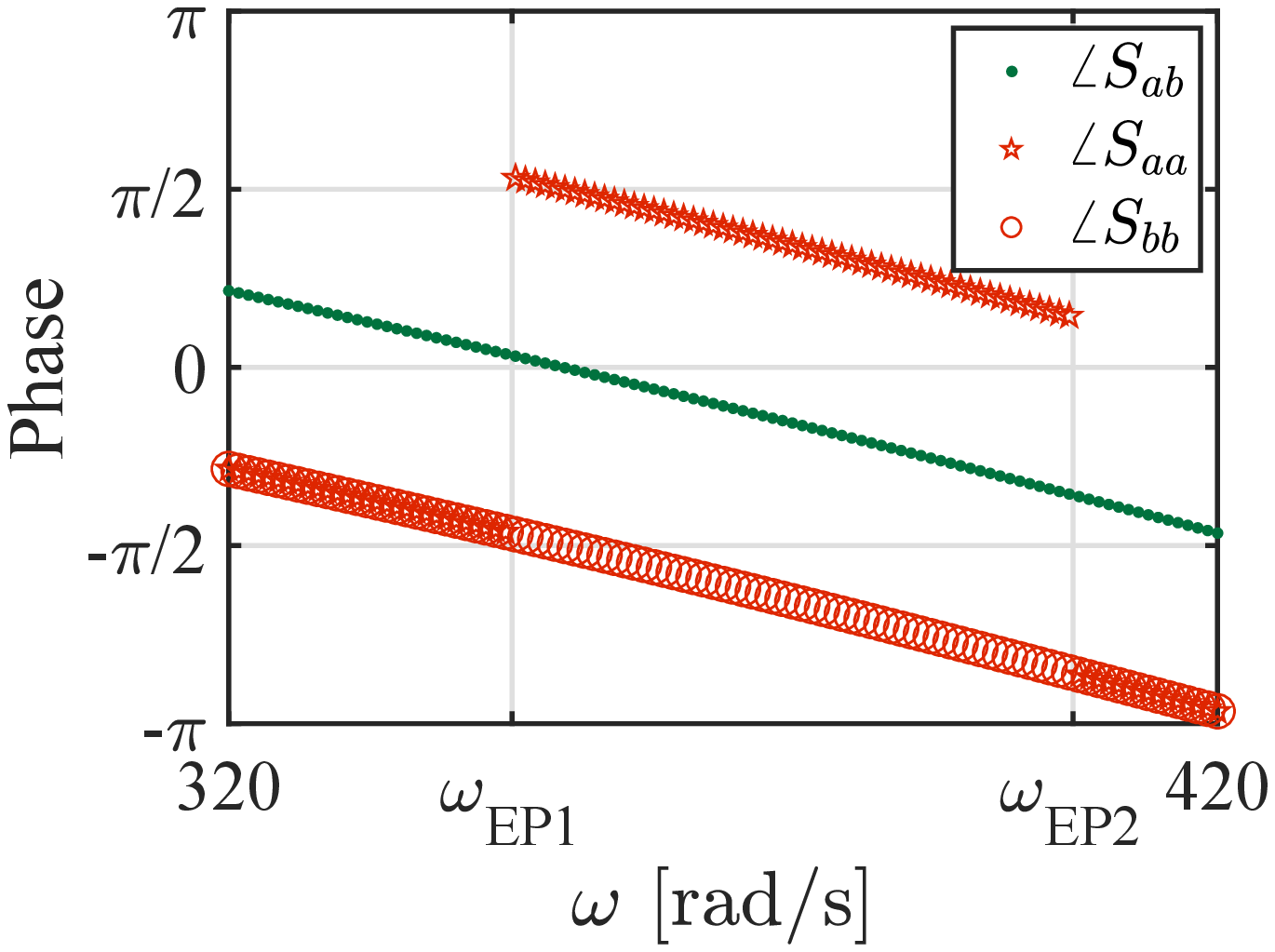}
		\caption{\label{fig:saphs}}
	\end{subfigure}%
	\begin{subfigure}[b]{0.4385\linewidth}
		\centering\includegraphics[height=140pt]{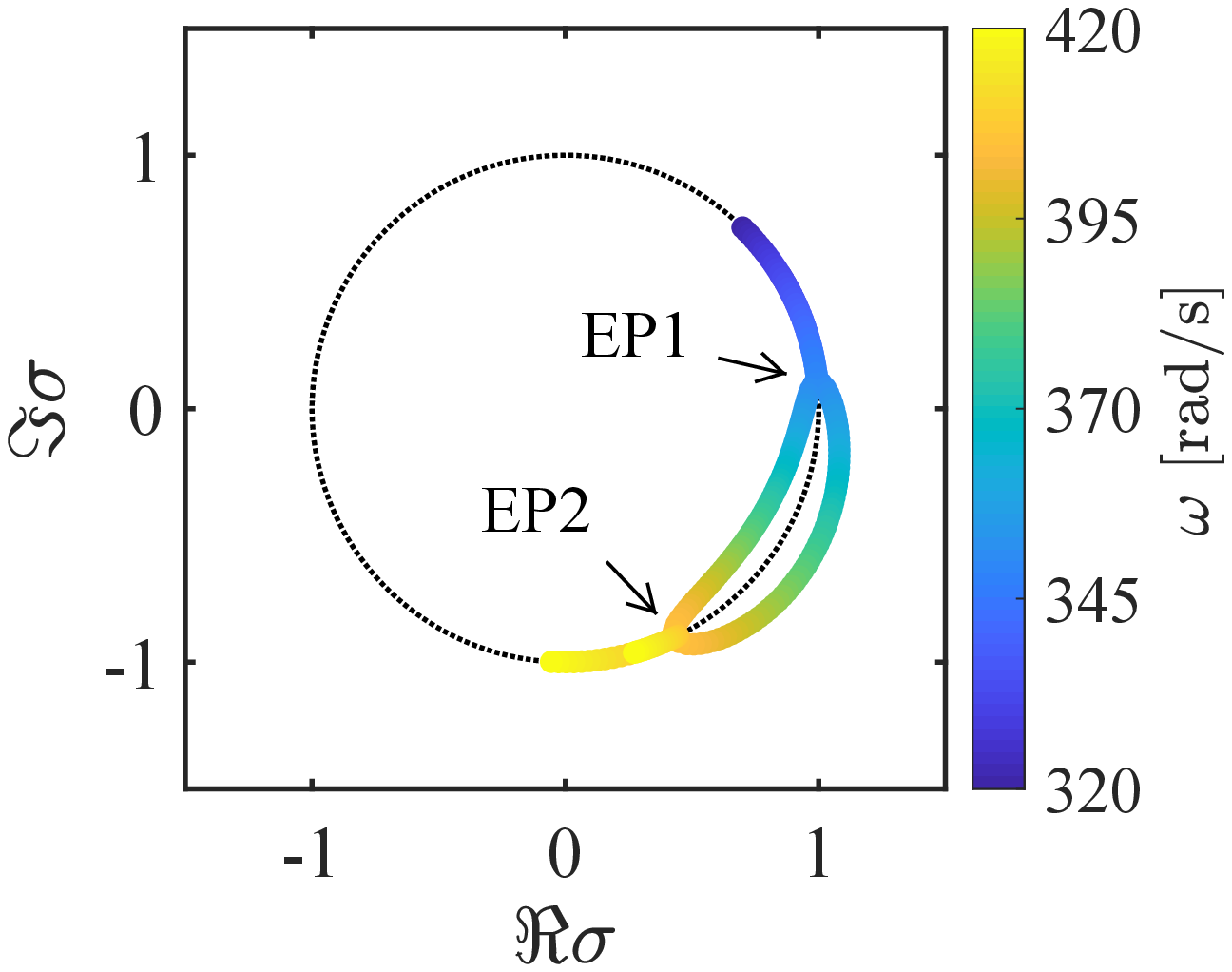}
		\caption{\label{fig:ssig}}
	\end{subfigure}
	\begin{subfigure}[b]{0.89\linewidth}
		\centering\includegraphics[height=140pt]{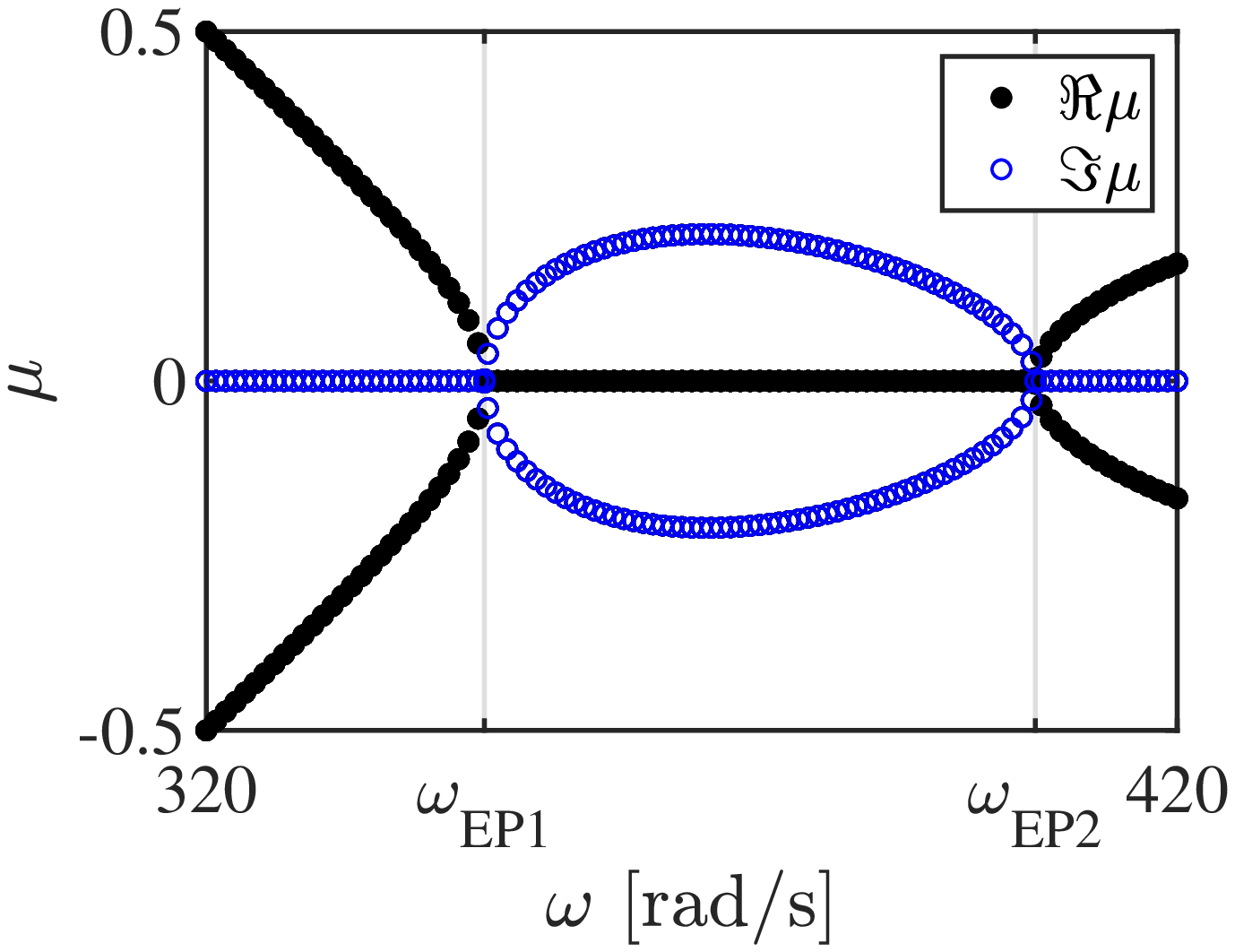}
		\caption{\label{fig:SEigVec}}
	\end{subfigure}%
	\caption{Scattering response near two EPs of the scattering matrix for a $\pt$ symmetric system. (\subref{fig:slam}) Amplitudes of the scattering coefficients in logarithmic scale.~(\subref{fig:stram}) Amplitudes of the transmission coefficients in linear scale for clarification.~(\subref{fig:saphs}) the scattering coefficient phases, ~(\subref{fig:ssig}) the eigenvalues of~$\TSM$ matrix, and~(\subref{fig:SEigVec}) the second components of the eigenvectors~\cref{eq:mu} of~$\TSM$ matrix. \label{fig:SMEP}} 
\end{figure}

The scattering responses are calculated and shown in \cref{fig:SMEP}, where the frequency range is~$[320,\ 420]\ \mathrm{rad/s}$. \Cref{fig:slam} shows the amplitudes of scattering coefficients in the logarithmic scale, where two poles of~$\SM_{aa}$ can be found at~$\omega_\mathrm{EP1}\approx\SI{348.7}{rad/s}$ and~$\omega_\mathrm{EP2}\approx\SI{405.4}{rad/s}$.
At these two frequencies, the left reflection coefficient~$\SM_{aa}$ becomes essentially zero, indicating one-way reflection. Based on \cref{eq:seiga,eq:seige}, the~$\TSM$ matrix exhibits coalescing eigenvalues and eigenvectors at~$\omega_\mathrm{EP(1,2)}$, as shown in \cref{fig:ssig,fig:SEigVec}. The eigenvectors are re-normalized as~
\begin{equation}\label{eq:mu}
    \vect{\mu}=\begin{pmatrix}1 \\ \mu\end{pmatrix}, 
\end{equation}
and only the second component is shown in \cref{fig:SEigVec}. The two EPs are labelled as EP1 and EP2, and they correspond to the phase transition thresholds where the~$\pt$ symmetry of the eigenvectors is spontaneously broken. At these EPs, the transmission amplitude becomes one, as shown in \cref{fig:stram}. Both~$\sigma$ and~$\mu$ bifurcate at the EPs. For the frequency smaller than~$\omega_\mathrm{EP1}$ or larger than~$\omega_\mathrm{EP2}$, the system is in the~$\pt-$unbroken phase, where the two reflection coefficients have the same phase (see \cref{fig:saphs}) and the transmission coefficient has amplitude smaller than one. The non-degenerate eigenvalues are both unimodular but different in phase, and the eigenvectors are real. The frequency range~$(\omega_\mathrm{EP1},\omega_\mathrm{EP2})$ represents the~$\pt-$broken phase. In the~$\pt-$broken phase, the eigenvalues have same phases but inverse amplitudes, i.e.,~$|\sigma_1\sigma_2|=1$. The second component of the eigenvector becomes purely imaginary. The reflection coefficients have exactly~$\pi$ difference in their phases. The transmission amplitude exceeds one, and all single-sided incident waves will be amplified. This single-sided reflection is most easily observed in the the transmission and reflection amplitudes shown in \cref{fig:slam}.

Due to the defectiveness of the scattering matrix, prescribed or measured states at~$\omega_\mathrm{EP(1,2)}$ can not be decomposed into the eigenvectors of ~$\TSM$.
On the other hand, a scattering state can always be decomposed into the two eigen-modes when operating in the~$\pt-$broken or unbroken phases. In the symmetry-unbroken phase, the two basis modes have purely real components and are invariant under~$\pt$ reversal. In the symmetry-broken phase, the two basis vectors no longer maintain the symmetry due to the imaginary components. Nevertheless, the~$\pt$ symmetry conditions in~\cref{table:1} are always satisfied.

\section{Conclusions}
Using a simple tunable discrete model setup, the exceptional points (EP) of the dynamic and scattering matrices of monatomic, diatomic, or locally resonant mechanical systems are analyzed. It is shown that the eigenfrequency band structure of a micro-structured medium can possess EPs as complex singularities or defects of the associated linear operators. Various phenomena associated with wave propagation in mechanical materials can be categorized and analyzed with this tool set. To summarize, the highlights of this work are:
\begin{itemize}
    \item Elucidation of EP-related phenomena such as level repulsion, mode coalescence, mode switching and self-orthogonality in a simple yet physical setup,
    \item Summary of the transfer and scattering matrix properties for general 1D (discrete and continuous) systems,
    \item Demonstration of unique scattering behavior at the EPs of the dynamic and scattering matrices (bi-directional reflectionless and single-sided reflection, respectively).
\end{itemize}

This study of discrete metamaterial systems contributes to fundamental understanding of the EPs in mechanical micro-structured media, and will be of interest to novel applications such as robust sensing and filtering. 
The complex valued springs (especially the ones with gains) used in this paper represent some practical challenges. However, they are helpful in the theoretical investigation of the topological and spectral properties of EPs, and ideas for their realization are already presented in literature. 
Furthermore, the discrete modeling approach can be utilized for conceptual design of novel devices as well as transferring the knowledge from EM and photonics domain into mechanical counterpart devices. See for example, the potential lasing mechanism~\cite{Zhang2019d} and prototype EP-enabled lasing devices studied in optics~\cite{Peng2016a}.
\newline

\noindent\textbf{ACKNOWLEDGEMENTS}
\newline
The authors acknowledge NSF grant \#1825969 to the University of Massachusetts, Lowell.
\newline

\noindent\textbf{DATA AVAILABILITY}
\newline
The data that support the findings of this study are available from the corresponding author
upon reasonable request.

\noindent\textbf{ACKNOWLEDGEMENTS}
\newline
The Version of Record of this article is published in The European Physical Journal Plus, and is available online at https://doi.org/10.1140/epjp/s13360-022-02626-6
\newline

%\appendix
\begin{appendices}
\section{Equations of motion derivation}\label{sec:appEOM}
\begin{figure}[!ht]
	\centering\includegraphics[width=180pt]{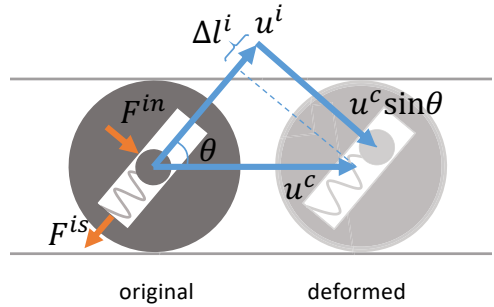}
	\caption{\label{fig:appEOM} Illustration of the forces and displacement of the local resonator. The subscript~$n$ is omitted here.}
\end{figure}

To derive the equations of motion (EOMs) of a unit cell in~\cref{DMEP}, an illustration is shown in \cref{fig:appEOM}. The cell springs~$\beta^c$ and neighbor cells are not shown here.
In one unit cell, the forces acting on the internal resonator include two components. The spring force~$F^{is}$ is parallel to the~$u^i$ direction, and is defined positive if spring is in tension. The rigid crystal mass wall applies a force~$F^{in}$ normal to~$u^i$ (parallel to component~$u^c\sin\theta$) since all contacting surfaces are frictionless. The crystal mass is kept from vertical motion by frictionless walls above and below it and therefore that DOFs does not enter the kinematics or dynamics equations. The net length increased in the resonator spring is~$\Delta l^i=u^i-u^c\cos\theta=u^i-\kappa u^c$. The tensile spring force is
\begin{equation}\label{eq:fis}
    F^{is}=\beta^i\Delta l^i=\beta^i(u^i_n-\kappa u^c_n).
\end{equation}
The EOM of the~$u^i$ DOF is
\begin{equation}
    M^i\frac{\partial^2u^i_n}{\partial t^2}=-F^{is}=\beta^i(\kappa u^c_n-u^i_n),
\end{equation}
The resonator also has a dependent DOF normal to~$u^i$, which is simply~$u^c \sin\theta$. The acceleration in this direction is caused by the wall force~$F^{in}$. Therefore, we have
\begin{equation}\label{eq:fin}
    M^i\sin\theta\frac{\partial^2u^c_n}{\partial t^2}=F^{in}.
\end{equation}
For the cell mass~$M^c$, its motion is allowed only in the horizontal direction. Therefore, its EOM is
\begin{equation}\label{eq:cellEOM1}
    M^c\frac{\partial^2u^c_n}{\partial t^2}=\beta^c(u^c_{n+1}-2u^c_n+u^c_{n-1})+F^{is}\cos\theta-F^{in}\sin\theta.
\end{equation}
The first term on the R.H.S. of~\cref{eq:cellEOM1} is the force applied by neighbor cells (not shown in~\cref{fig:appEOM}). The second and third terms are the forces supplied by the resonator, projected onto the horizontal direction. Substituting~\cref{eq:fis} and~\cref{eq:fin} into~\cref{eq:cellEOM1} yields
\begin{equation}
    M^c\frac{\partial^2u^c_n}{\partial t^2}=\beta^c(u^c_{n+1}-2u^c_n+u^c_{n-1})+\kappa\beta^i(u^i_n-\kappa u^c_n)-(1-\kappa^2) M^i\frac{\partial^2u^c_n}{\partial t^2},
\end{equation}
where~$\kappa=\cos\theta$ and~$1-\kappa^2=\sin^2\theta$.

\section{Unit cell transfer matrix}\label{appTM}

At the boundaries of the cell, the state vectors are: 
\begin{equation}
\vect{\psi}^{l,r}=\begin{pmatrix}v^{l,r} \\  N^{l,r}\end{pmatrix},
\end{equation} 
where~$l$ or $r$ denotes left or right,~$v^{l,r}=\mathrm{i}\omega u^{l,r}$ is the particle velocity in~$x$ direction, and~$N^{l,r}$ is the internal normal traction force in the springs applied at the boundary (tensile positive, relating to normal stress component in a continuum system). The spring constitutive equations are:
\begin{align}
N^l&=2\beta^p(u^c-u^l),\label{TL}\\
N^r&=2\beta^q(u^r-u^c)\label{TR}.
\end{align}
The equation of motion for the main crystal chain mass DOF is:
\begin{equation}
N^r-N^l+\kappa\beta^i(u^i-\kappa u^c)+\omega^2M^{ci}u^c=0,\label{TC}
\end{equation}
where~$\kappa$ and~$M^{ci}$ are quantities defined in the main text.
The equation of motion for the internal resonator is:
\begin{equation}
\beta^i(u^i-\kappa u^c)=\omega^2 M^{i}u^i.\label{TI}
\end{equation}
Combining \crefrange{TC}{TI}, the crystal displacement can be written as:
\begin{equation}
u^c=\frac{N^l-N^r}{K_T},\label{eq:uckt}
\end{equation}
where
\begin{equation}
K_T=\dfrac{\kappa^2 M^i \omega^2}{1-(\omega/\omega^i)^2}+M^{ci}\omega^2,
\end{equation}
which would also simplify to $K_T = M^c \omega^2$ in the limit when $M^i = 0$. 
Substituting \cref{eq:uckt} into \crefrange{TL}{TR}, the state vectors at the boundaries of a unit cell can be written as:
\begin{equation}
\label{eq:cellTM}
\begin{pmatrix}
v^r\\N^r\end{pmatrix}=
\TM^{cell}
\begin{pmatrix}
v^l\\N^l\end{pmatrix},
\end{equation} 
with
\begin{equation}\label{eq:TMRUC}
\TM^{cell}=
\begin{pmatrix}
	1-\dfrac{K_T}{2\beta^q} & \mathrm{i}\omega\dfrac{2\beta^p+2\beta^q-K_T}{4\beta^p\beta^q}  \\
	\dfrac{\mathrm{i}K_T}{\omega} & 1-\dfrac{K_T}{2\beta^p}
\end{pmatrix}~.
\end{equation}
The eigenvectors of a non-defective transfer matrix span~$\mathbb{C}^2$ and therefore form a basis for any possible state. Thus any state vector $\vect{\psi}$ observed can be decomposed into a superposition of two eigenmodes, i.e., one can identify and separate (any wave or any linear combination of) the forward and backward wave components in this 1D case into eigenmodes of the transfer matrix of the finite specimen. 

\section{Scattering matrix}\label{appSM}

%Note that the ordering of the scattering parameters in $\TSM$ is different than traditional ordering of $\SM$ in which the order of the ports ($a$ or $b$) are kept same in both incoming and outgoing columns. In the present representation, the components are ordered based on right and left traveling nature in the exterior domains. The reason for this choice will become apparent later.

The state vectors at locations~$x^a$ and~$x^b$ are simply derived using superposition:
\begin{equation}\label{eq:yab}
\vect{\psi}^{a,b}=\ii \omega e^{\ii \omega t}\begin{pmatrix}
1&1\\-Z^{a,b} &Z^{a,b}
\end{pmatrix}\begin{pmatrix}
A^{(a,b)+}\\A^{(a,b)-}
\end{pmatrix},
\end{equation}
where the impedance for bar $a$ or $b$
is~$Z^{a,b}=-N^{(a,b)+}/v^{(a,b)+}=\pi r_0^2\sqrt{E_0\rho_0}$. For a continuum~\cite{Amirkhizi2017}, the impedance is simply~$Z=-\sigma/v=\sqrt{E_0 \rho_0}$. For the continuum-discrete interfaces here, one needs to include the bar cross-sectional area in calculation.
Since it was chosen that $x^{a,b} = x^{l,r}$ (associated with the left and right boundaries of the full system) then 
\begin{equation}
\label{eq:sampleTM}
\vect{\psi}^b=
\TM
\vect{\psi}^a=\begin{pmatrix}
\TM_{11} &\TM_{12}  \\
\TM_{21} &\TM_{22}
\end{pmatrix}
\vect{\psi}^a,
\end{equation}
due to the construction of~$\TM=\TM^{(j)} \cdots\TM^{(2)}\TM^{(1)}$ as the transfer matrix of entire sample (in total $J$ cells) from~$x^l$ to~$x^r$, where and~$\TM^{(j)}$ is the TM of~$j$-th cell counting from the left interface~$x^l$, and $\vect{\psi}^{a,b} = \vect{\psi}^{l,r}$.
Substituting \cref{eq:yab} into \cref{eq:sampleTM}, the scattering coefficients in \cref{eq:asa} can be obtained analytically: 
\begin{align}
\SM_{aa}&=\frac{-\TM_{21}+\TM_{22}Z^b-\TM_{11} Z^a+\TM_{12}Z^a Z^b}{\Delta},\label{eq:saa}\\
\SM_{bb}&=\frac{-\TM_{21}+\TM_{11}Z^a-\TM_{22}Z^b+\TM_{12}Z^a Z^b}{\Delta},\label{eq:sbb}\\
\SM_{ba}&=\frac{2Z^b}{\Delta}\abs{\TM},\label{eq:sba}\\
\SM_{ab}&=\frac{2Z^a}{\Delta},\label{eq:sab}\\
\Delta&=\TM_{21}+\TM_{11}Z^a+\TM_{22}Z^b+\TM_{12}Z^a Z^b,\label{eq:dels}
\end{align}

\end{appendices}

%
% For tables use
%\begin{table}
%\centering
%\caption{Please write your table caption here}
%\label{tab:1}       % Give a unique label
% For LaTeX tables use
%\begin{tabular}{lll}
%\hline\noalign{\smallskip}
%first & second & third  \\
%\noalign{\smallskip}\hline\noalign{\smallskip}
%number & number & number \\
%number & number & number \\
%\noalign{\smallskip}\hline
%\end{tabular}
% Or use
%\vspace*{5cm}  % with the correct table height
%\end{table}

\normalem
\bibliographystyle{aipnum4-1}
\bibliography{Manuscript}

\end{document}